\documentclass{aa}

\usepackage{color}
\usepackage{natbib}
\usepackage{graphicx}
\usepackage{caption}
\usepackage{subcaption}
\usepackage{mathabx}
\usepackage{txfonts}

\begin{document}

   \title{Two-population model of type Ia supernovae and their associations with host galaxies in ZTF DR2}
   \titlerunning{Two-population model of type Ia supernovae and their associations with host galaxies in ZTF DR2}
   
   \author{Rados{\l}aw Wojtak\thanks{Corresponding author: radek.wojtak@nbi.ku.dk}
          \and Lucas Hallgren
          \and Jens Hjorth
          }

   \institute{DARK, Niels Bohr Institute, University of Copenhagen, Jagtvej 155, 2200 Copenhagen, Denmark
                          }

   \date{}

  \abstract
   {}
    {
    We constrain type Ia supernova intrinsic properties, extinction, and probabilistic supernova–host associations using the volume-limited sample from the Zwicky Transient Factory DR2, 
  the largest selection-free data set of type Ia supernovae to date.
        }
   {
   We employ Bayesian hierarchical modelling to jointly analyse the distribution of SALT2 light-curve parameters and global host-galaxy properties 
   (stellar mass and rest-frame $g-z$ colour). The adopted model is a mixture of distributions representing two supernova populations corresponding to two distinct 
   modes of the stretch-parameter distribution, and two host-galaxy populations corresponding to two modes in the host-galaxy parameter space (red/massive and blue/less massive). 
   Motivated by observations, high-stretch supernovae are allowed to populate both host-galaxy populations, whereas low-stretch supernovae are assumed to be exclusively associated 
   with the red/massive host-galaxy population.
   }
   {
   The apparent non-linearity of the supernova magnitude–stretch relation implies a luminosity gap between the supernova populations, with the low-stretch population being 
   $\Delta M_{\rm B}=0.14\pm0.03$ mag brighter at $x_{1}=0$, and different slopes ($\Delta\alpha=0.065\pm0.024$, steeper for the low-stretch population). The mean extinction coefficient 
   is $R_{\rm B}=3.89\pm0.29$ (consistent with typical Milky Way values) in the blue/less massive host-galaxy population, which contains 68 per cent of high-stretch supernovae, and 
   $R_{\rm B}=3.08\pm0.08$ in the red/massive host-galaxy population. The difference in the mean $R_{\rm B}$ between the populations, together with a flatter slope of the 
   intrinsic-colour correction for blue colours ($\beta=2.06\pm0.45<R_{\rm B}$), represents the main non-linear effects in the supernova magnitude–colour relation 
apparent in the data. Host-galaxy step corrections, both in stellar mass and colour, naturally emerge from the way the two supernova populations, characterised by different intrinsic 
luminosities and extinctions, are distributed across host-galaxy populations.
      }
   {The high-stretch supernova population appears to be standardised with higher precision than its low-stretch counterpart ($0.06$~mag versus $0.11$~mag residual scatter). The adopted framework for modelling supernova populations and supernova-host associations makes host-galaxy step corrections redundant.
    }


   \maketitle

\section{Introduction}

Cosmologically useful distance measurements derived from type Ia supernova observations rely on applying a range of corrections 
collectively known as type Ia supernova standardisation. The most minimalistic, first-order approach corrects for two well-established empirical 
relations: redder--dimmer and wider light curve--brighter \citep{Phillips1999}. Commonly referred to as the Tripp calibration \citep{Tripp1998}, this standardisation 
model reduces the systematic error in distance moduli per supernova (the scatter in Hubble residuals) to $0.10-0.15$~mag, which is a factor of approximately 3 
smaller than that obtained without any standardisation. This enhancement in precision was one of the key elements that enabled the discovery of cosmic 
acceleration \citep{Riess1998,Perlmutter1999}. The Tripp calibration model is currently considered the main point of reference for developing its augmentations, 
motivated by both theory and observations, that attempt to provide a more complete explanation of remaining Hubble residuals.

The most popular strategy for improving the Tripp calibration is to account for possible correlations with host-galaxy properties. Initially discovered as 
a trend for more luminous supernovae to occur in massive ($M_{\star}\gtrsim 10^{10}M_{\odot}$) galaxies \citep{Kelly2010,Sullivan2010}, the relation 
was adopted as an additional correction applied to Tripp-corrected magnitudes in the form of a step function of host-galaxy stellar mass. Similar host-galaxy step corrections 
(sometimes modulated by supernova colour) were found for a range of mutually correlated variables, including rest-frame colours \citep{Roman2018}, 
stellar age \citep{Wiseman2022}, local specific star formation rate \citep{Rigault2020}, metallicity \citep{Padilla2026} and morphological class \citep{Pruzhinskaya2020}, 
with the largest step amplitude found for the local specific star formation rate \citep{Briday2022}. Their dependence on supernova reddening \citep{Brout2021,Vincenzi2024}, together with a reduction of the amplitude in the outer regions of their host galaxies \citep{Toy2025}, points to an important role of extinction and its variation across host-galaxy types. 
Host-galaxy step corrections are most likely manifestations 
of two distinct supernova populations, whose origins are correlated with their host galaxies. 
These two populations are expected to have different 
luminosities (with possible modulation by extinction), such that the empirical step functions can arise from mixing these populations in host-galaxy parameter space. Disentangling these two populations in the space of supernova intrinsic variables (e.g. luminosity, intrinsic colour) is key to understanding the physical origin of host-galaxy step corrections, especially their achromatic (extinction-independent) component.

A significant step towards understanding the residual scatter in Tripp-corrected magnitudes was enabled by Bayesian hierarchical modelling, applied 
directly to supernova light curves \citep{Mandel2022} or to their compressed representations in the form of light curve parameters \citep[see e.g.][]{Wojtak2023}. 
In this approach, one uses the full information contained in the distributions of observable to constrain the statistical properties of latent 
variables and relations between them, often reflected by non-linear, and generally probabilistic relations between observables. A prime example is disentangling 
supernova intrinsic colour from host-galaxy reddening, thereby enabling estimates of the total-to-selective extinction parameter.

Several recent studies have incorporated the concept of two supernova populations as a mixture of probability densities in Bayesian hierarchical modelling 
of type Ia supernova data \citep{Wojtak2023,Wojtak2025,Rubin2026}. The two populations emerge as two modes in the distribution of light-curve widths, 
clearly visible at low redshifts across various supernova compilations \citep{Scolnic2018,Ginolin2024a}. These two modes have long been associated with generically young 
(longer light curves or high "stretch") and old (shorter light curves or low "stretch") progenitors, as governed by host-galaxy age \citep{Sullivan2006} or progenitor age 
\citep{Kim2026,Raju2026}. Recent modelling by \citet{Wojtak2023,Wojtak2025,Rubin2026} points to substantial differences between the supernova populations represented by the two modes in the stretch distribution, 
both in terms of intrinsic properties including average luminosities, and extinction. Two-population 
models generally predict non-linear corrections to the traditional stretch correction of the Tripp calibration as a consequence of population mixing 
in the latent variable space. These corrections correlate closely with the relative fraction of massive host galaxies \citep{Wojtak2025}, suggesting that the mass step 
can be interpreted as an emergent property within this framework. In fact, the mass step vanishes when it is constrained simultaneously with the two supernova 
populations \citep{Rubin2026}.

Modelling two supernova populations is also relevant for obtaining more reliable constraints on host-galaxy extinction. Supernova data exhibit a wide 
range of possible total-to-selective extinction coefficients, reflected in the scatter of the effective slopes of the magnitudes-colour relation in the regime of 
red colours \citep[see e.g.][]{Gonzalez2021,Brout2021}. Various approaches based on modelling either the supernova populations directly or their proxies 
given by correlated host-galaxy properties point to a Milky-Way-like \citep{Fitzpatrick2007,Schlafly2016} average extinction in the population of high-stretch supernovae 
\citep{Wojtak2023,Wojtak2025} or in associated late-type host galaxies \citep{Vincenzi2024,Popovic2021,Rubin2026}. On the other hand, substantially lower effective extinction coefficient, 
which is not observed in the Milky Way and other galaxies \citep[][]{Salim2018,Duarte2023}, is found for low-stretch supernovae 
or their associated early-type hosts.

Following recent informative constraints on type Ia supernova populations associated with the modes of the stretch distribution, obtained from Pantheon+ and Union3.1 supernova compilations \citep{Wojtak2023,Wojtak2025,Rubin2026}, 
we apply the two-population model to type Ia supernova light curve parameters from the Zwicky Transient Factory (ZTF) DR2 \citep{Rigault2025}. Motivated by 
observed correlations between type Ia supernova stretch and host-galaxy properties including stellar mass and specific star formation rate \citep{Sullivan2010,Rigault2020,Smith2020,Popovic2021a}, 
we augment the framework with probabilistic modelling of associations between the supernova populations and the two primary populations of host galaxies imprinted in the joint distribution 
of global stellar masses and rest-frame $(g-z)$ colours. Including the host-galaxy information helps separate the supernova populations in regions of light curve parameter space where they strongly 
overlap (e.g. near the mean stretch). We model both supernova and host-galaxy data in a joint analysis, accounting for all known sources 
of uncertainties and systematic effects including the so-called pocket effect \citep{Rigault2025}. The ZTF DR2 data set is the largest volume-limited sample of spectroscopically confirmed 
type Ia supernovae. Its unprecedented constraining power has been demonstrated in a series of studies showing, among other things, non-linearity in the stretch correction, 
the host-galaxy mass-step correction, the bimodality in the light curve stretch \citep{Ginolin2024a}, and an unambiguous fraction of highly reddened supernovae with apparent colours exceeding 
the upper limits typically applied to cosmological sample \citep{Ginolin2025}.

The outline of the paper is as follows. In Section~\ref{sec:data}, we describe the data, including the criteria used for selecting supernovae with well-measured light curve parameters, 
peculiar velocity corrections and related uncertainties, as well as the implementation of the pocket-effect correction. Section~\ref{sec:model} outlines the two-population model and 
the assumptions underlying the adopted priors for supernova and host-galaxy variables. The results of fitting the model to the data are presented in 
Section~\ref{sec:results} and discussed in Section~\ref{sec:discussion}. We summarise our findings and conclude in Section~\ref{sec:summary}.

\section{Data}
\label{sec:data}
We used measurements of SALT2 \citep{Guy2007} light curve parameters of type Ia supernovae 
from the ZTF DR2 catalogue \citep{Rigault2025}\footnote{https://ztfcosmo.in2p3.fr}. The parameters are the dimensionless flux amplitude 
$x_{0}$, the stretch parameter $x_{1}$ and the colour parameter $c$, which is defined such that it closely 
approximates the rest-frame $B-V$ colour at the rest-frame $B$-band peak \citep[with a precision of about $0.01$~mag;][]{Kessler2013}. 
We converted the dimensionless flux amplitude $x_{0}$ to the corresponding apparent 
rest-frame $B$-band magnitude $m_{\rm B}$ using
\begin{equation}
m_{\rm B,ZTF}=-2.5\log_{10}(x_{0})+10.635
\end{equation}
\citep[see e.g.][]{Kessler2013}. The magnitude--flux relation was also used to transform all covariance elements 
related to $x_{0}$.

We transformed heliocentric redshifts $z_{\rm hel}$ listed in the ZTF DR2 catalogue to the CMB rest frame using the 
Planck dipole measurement \citep{Planck2020}. The resulting redshifts $z_{\rm cmb}$ were corrected for peculiar 
velocities using the cosmic flow model \citep{Carrick2015}\footnote{https://cosmicflows.iap.fr}. The corrected redshift 
$z_{\rm cos}$ was computed by solving the following set of equations
\begin{eqnarray}
(1+z_{\rm cmb}) & = & (1+z_{\rm cos})[1+v_{\rm pec}(R_{\rm SN},\alpha_{\rm SN},\delta_{\rm SN})/c] \nonumber \\
H_{0}R_{\rm SN} & = & c\Big(z_{\rm cos}-\frac{1+q_{0}}{2}z_{\rm cos}^{2}\Big),
\end{eqnarray}
where $v_{\rm pec}$ is the radial component of the peculiar velocity at the supernova 3D position in the model's box, $R_{\rm SN}$ 
is the comoving distance to the supernova, $\{\alpha_{\rm SN},\delta_{\rm SN}\}$ are supernova celestial coordinates, $q_{0}=-0.53$ is the deceleration parameter \citep{Planck2020}, and $H_{0}=100h$~km~s$^{-1}$~Mpc$^{-1}$. 
The corrections were applied only to those supernovae that are within the spatial boundaries of the model. The maximum comoving distance 
with available constraints on the peculiar velocity is about $200$~Mpc~h$^{-1}$, and it coincides with the size of the volume-limited 
sample ($z_{\rm cos}<0.06$). The precision of the peculiar velocity model is typically assumed to be about 
$\sigma_{\rm pec}=200$~km~s$^{-1}$ and it accounts for residuals in the modelled velocity field arising 
primarily from its non-linear, and thus minimally correlated, components \citep{Carrick2015}. We propagated this uncertainty to the errors in distance 
moduli derived from redshifts $z_{\rm cos}$. The total error in the distance modulus is given by
\begin{equation}
\sigma_{\mu}=(5/\ln10)(\sigma_{\rm pec}^{2}/ (c\,z_{\rm cos})^{2}+\sigma_{z}^{2}/ z_{\rm cos}^{2})^{1/2},
\label{mu_error}
\end{equation}
where $\sigma_{z}$ is the measurement error listed in the ZTF DR2 catalogue. Including measurement errors 
is particularly relevant for redshifts estimated from supernova (as opposed to galaxy) spectra (21 per cent of the final sample selected for this study), for 
which $\sigma_{\mu}$ becomes comparable to the modelled signal. With a typical value of $3\times 10^{-3}$, the resulting error in 
distance modulus is $0.1$~mag at the maximum redshift of the volume-limited sample ($z_{\rm cos}=0.06$) and as large as $0.3$~mag 
at redshift $z_{\rm cos}=0.02$. For the former, the error is comparable to the scatter in supernova Hubble diagrams with the Tripp calibration, 
whereas for the latter it is comparable to the difference between colour corrections at two colours separated by $\Delta c\approx0.1$, which is comparable to a typical range of intrinsic colours.

\begin{figure*}
   \centering
   \includegraphics[width=0.95\linewidth]{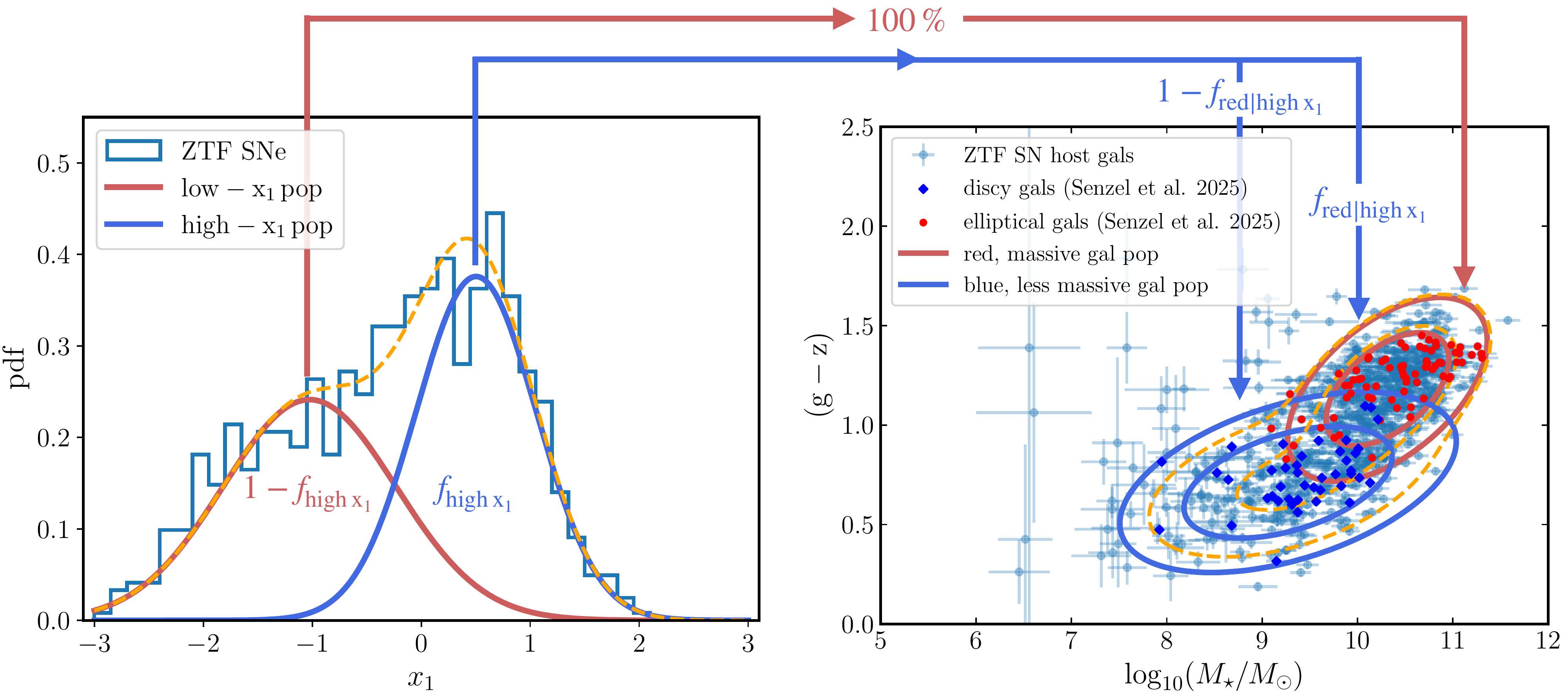}
    \caption{Supernova and host-galaxy populations observed as bimodalities in the stretch parameter distribution (\textit{left panel}) 
    and the joint distribution of the global host galaxy stellar masses and rest-frame $(g-z)$ colours (\textit{right  panel}). The light blue 
    data points and histogram show the data from the ZTF DR2 volume-limited supernova sample used in this study. The dark blue and red 
    lines show the Gaussian components of mixture models (orange dashed lines) fitted to the data. The directed graph represents the basic 
    structure of the prior probability used in our modelling in which the low-$x_{1}$ supernova population is fully associated with the red/massive 
    host galaxy population, whereas the high-$x_{1}$ supernovae are linked to either red or blue host galaxy populations.}
              \label{pops}
    \end{figure*}

We restricted our analysis to the volume-limited sample ($0.01<z_{\rm cos}<0.06$) of normal type Ia supernovae commonly used in cosmological 
analysis ($sntype=snia\_cosmo$, discarding the cases with undefined types). As in the main ZTF analyses of colour and stretch corrections 
\citep{Ginolin2024a,Ginolin2025}, we further restricted the sample to supernovae with good light curve sampling and basic fit-quality cuts as defined 
by \citet{Rigault2025}. The applied quality cuts include limits on the stretch parameter, $-3\le x_{1}\le 3$, and the colour parameter, $-0.2\le c\le 0.8$~mag. 
All these selection conditions result in a sample containing 891 supernovae. We further discarded supernovae with probability fits $p<0.001$, 
for which SALT2 best-fit models are not expected to adequately represent the observed light curves. These correspond to 66 cases, compared to about 1 expected for 
fits consistent with the $\chi^{2}$ distribution. Eliminating additional 16 supernovae for which host-galaxy photometry is missing or the 
global rest-frame $(g-z)$ lies outside the range $[0,2]$~mag, we obtained the final sample consisting of 809 supernovae.

We used the global stellar masses and rest-frame $g-z$ colours of host galaxies to model connections between type Ia supernova populations 
and their host galaxies. Stellar mass uncertainties listed in the ZTF DR2 catalogue do not include the contribution from $\sigma_{\mu}$. In order 
to obtain more realistic uncertainties (especially at small redshifts), we added $0.4\sigma_{\mu}$ given by eq.~(\ref{mu_error}) in quadrature. 
All joint fits to supernova and host-galaxy data in our analysis neglected the resulting partial correlation between stellar mass 
and the supernova distance modulus.

The ZTF observations after November 2019 were affected by the so-called pocket effect, which is a non-linearity in the ZTF CCD readout 
\citep{Rigault2025}. The effect impacts flux measurements via alterations to the point-spread function. It is of the order of 1 per cent between
15~mag and 19~mag, and it is independent of colour. In a first-order approximation, the effect shifts the estimates of the stretch parameter $x_{1}$ 
towards smaller values (more rapidly declining light curves) and modifies peak magnitudes as a function of the true flux \citep{Rigault2025}. 
Although both effects appear to have a negligible impact on modelling supernova Hubble residuals as a function of $x_{1}$ and $c$ 
\citep{Ginolin2024a,Ginolin2025}, we account for them by applying direct corrections to the measured stretch parameter and peak 
magnitude. For the former, we assumed that unbiased measurements of $x_{1}$ are given by
\begin{equation}
x_{1}=x_{1,{\rm ZTF}}+\Delta_{x_{1}}\theta(t_{0}-58757.5{\rm MJD}),
\end{equation}
where $x_{1,{\rm ZTF}}$ is the stretch parameter measurement listed in the ZTF DR2 catalogue (affected by the pocket effect), 
$\Delta_{x_{1}}$ is a nuisance parameter, $\theta(x)$ is the Heaviside step function ($\theta(x>0)=1$ and $\theta(x<0)=0$), $t_{0}$ 
is the peak magnitude epoch and 58757.5~MJD is a date of transition between observation periods unaffected and affected by 
the pocket effect (chosen as 1 October 2019). For the latter, we assumed a linear relation between the true and observed magnitudes. 
Although the pocket effect is in general non-linear, the linear approximation appears to be sufficient between about 16~mag 
and 19~mag \citep[see Fig. 2 of][]{Rigault2025}, which corresponds to the range containing the majority (about 93 per cent) of the supernova 
peak magnitudes in the volume-limited sample. The linear model is also consistent with the assumed stretch correction given by a global shift, which would otherwise become 
dependent on the peak magnitude. The peak magnitudes $m_{\rm B}$ corrected for the pocket effect are modelled as
\begin{equation}
m_{\rm B}=m_{\rm B,ZTF}-a_{\rm pocket}(m_{\rm B,ZTF}-17.2)\theta(t_{0}-58757.5{\rm MJD}),
\end{equation}
where $a_{\rm pocket}$ is a nuisance parameter and 17.2~mag is a reference magnitude at which the pocket effect vanishes \citep[see Fig. 2 of][]{Rigault2025}. 
From modelling the pocket effect and its impact on light curve parameter estimation, one can expect $a_{\rm pocket}\approx 0.015$ 
\citep[estimated from Fig. 2 of ][]{Rigault2025} and $\Delta_{x_{1}}\approx 0.1$ \citep{Rigault2025,Ginolin2024a}. However, instead of fixing these 
parameters, we fitted them simultaneously with the main parameters describing supernova and host-galaxy populations.

\section{Model}
\label{sec:model}
We modelled the joint distribution of supernova light curve parameters $\xi_{\rm SN}=\{m_{\rm B},x_{1},c\}$ and the global 
host-galaxy properties $\xi_{\rm host}=\{\log_{10}(M_{\star}/M_{\odot}),(g-z)\}$. Hereafter, we refer to the joint supernova-host observable 
as $\xi=\{\xi_{\rm SN},\xi_{\rm host}\}$. We adopted a commonly used model assuming linear relations between the supernova 
peak magnitudes, the stretch parameter $x_{1}$, the intrinsic colour $c_{\rm int}$ and the host-galaxy reddening $E(B-V)$,
\begin{eqnarray}
m_{\rm B} & = & M_{\rm B}+\mu(z_{\rm cos})-\alpha x_{1}+\beta c_{\rm int}+R_{\rm B}E(B-V) \nonumber \\
c & = & c_{\rm int}+E(B-V),
\label{obs-lat_eq}
\end{eqnarray}
where $\mu(z_{\rm cos})$ is the distance modulus \citep[assuming the Planck cosmology,][with the Hubble constant renormalised 
to $H_{0}=70\,{\rm km}\,{\rm s}^{-1}\,{\rm Mpc}^{-1}$]{Planck2020_cosmo}, $R_{\rm B}$ is the effective total-to-selective extinction coefficient, and $M_{\rm B}$ is the 
supernova absolute magnitude. Using Bayesian hierarchical modelling, we placed constraints on the distributions of all latent variables 
$\phi_{\rm SN}=\{M_{\rm B},\alpha,\beta,c_{\rm int},x_{1},E(B-V),R_{\rm B}\}$ governing the observed light curve parameters. 
Degeneracies between some latent variables apparent in eq.~(\ref{obs-lat_eq}), notably between $c_{\rm int}$ and $E(B-V)$),
are broken as a result of differences between prior distributions. We assumed Dirac delta functions for $\alpha$ and $\beta$, Gaussian prior 
distributions for $M_{\rm B}$, $c_{\rm int}$, $x_{1}$ and $R_{\rm B}$, and an exponential model for the reddening $E(B-V)$, i.e.
\begin{equation}
p_{E(B-V)}(E(B-V))=\frac{1}{\tau}\exp\Big(-\frac{E(B-V)}{\tau}\Big).
\end{equation}
The exponential model is commonly used in this type of modelling. The prior can be motivated as the maximum-entropy 
solution for a positively defined variable ($E(B-V)>0$) given a known mean ($\langle E(B-V)\rangle=\tau$). The exponential 
model is also thought to approximate the actual host-galaxy reddening distribution in discy galaxies. However, recent studies 
that simulate the reddening effect in relation to supernova positions and dust distribution have revealed noticeable 
deviations from the exponential model \citep{Hallgren2025,Duarte2026}. We address this issue in Section~\ref{sec:discussion}.

\subsection{Supernova and host galaxy populations}

The first key assumption of our model is that type Ia supernovae originate from two distinct populations whose primary 
observational manifestation is the bimodality of the stretch parameter distribution. The bimodality is a well-confirmed 
property in various supernova samples at small redshifts \citep[see e.g.][]{Scolnic2018,Wojtak2023,Wojtak2025}, 
including the ZTF volume-limited sample \citep{Ginolin2025}, and at high redshifts, although with an evolutionarily 
suppressed low-stretch component \citep{Nicolas2021,Rubin2026}.

The left panel of Figure~\ref{pops} shows the stretch parameter distribution for the supernovae used in this study, as well as the best-fit model 
of a mixture of two Gaussian distributions representing the two supernova populations, hereafter referred to as high-$x_{1}$ 
and low-$x_{1}$ populations. In the final model applied to the complete supernova data, we assumed two independent sets of 
prior distributions related to $\phi_{\rm SN}$ (the same prior parameterisations but different hyperparameters) for both populations. 
Due to the lack of sufficient constraining power, we kept $\beta$ as the only parameter common to both populations 
(see also Section~\ref{sec:res_int}).

We separated the host galaxies into two populations represented by two bivariate normal distributions fitted to the global stellar masses 
and rest-frame $(g-z)$ colours. The model distributions include also a uniform background accounting for possible outliers 
(see the right panel of Fig.~\ref{pops} and eq.~(\ref{prior_host}) below). The two galaxy populations correspond closely to early-type galaxies (red, massive) and late-type 
galaxies (blue, less massive), although the former is also expected to overlap with massive, star-forming galaxies reddened by dust \citep{Ramaiya2025}. 
Hereafter, we refer to them as red and blue host galaxy populations. The right panel of Figure~\ref{pops} shows 
the contours representing the distribution of galaxies in both populations. The figure also demonstrates that elliptical and disc galaxies 
classified by \citet{Senzel2025} match well the red and blue populations, respectively. The classification was based on fitting the light 
profiles and decomposing them into bulge and disc components, and it is available for only a fraction of all host galaxies from ZTF DR2.

The second key assumption of our model concerns the connection between the supernova and host galaxy populations. 
Observations show that low-stretch supernovae are strongly associated with early-type galaxies, whereas the high-stretch ones 
are found in both late- and early-type galaxies, with a preference for the former \citep[see e.g. Fig.~1 of][]{Ginolin2025}. 
Motivated by these findings, we assumed that the low-stretch supernova population is associated only with the red host galaxy population, whereas the high-stretch supernovae 
are partially linked to blue host galaxies and partially to red host galaxies. Figure~\ref{pops} shows an intuitive schematic of these associations. 
In the joint space of light curve parameters and host galaxy parameters, these relations are captured by the following prior probability distribution:
\begin{align}
&p_{\rm prior}(\pmb{\phi}) =(1-f_{\rm high\,x_{1}})p_{\rm SN}(\pmb{\phi}_{\rm SN}|\pmb{\Theta}_{\rm SN,low\,x_{1}})p_{\rm host}(\pmb{\phi}_{\rm host}|\pmb{\Theta}_{\rm host,red})+ \nonumber \\
&f_{\rm high\,x_{1}}\Big(f_{\rm red|high\,x_{1}}p_{\rm SN}(\pmb{\phi}_{\rm SN}|\pmb{\Theta}_{\rm SN,high\,x_{1},red})p_{\rm host}(\pmb{\phi}_{\rm host}|\pmb{\Theta}_{\rm host,red})+\nonumber \\
&(1-f_{\rm red|high\,x_{1}})p_{\rm SN}(\pmb{\phi}_{\rm SN}|\pmb{\Theta}_{\rm SN,high\,x_{1},blue})p_{\rm host}(\pmb{\phi}_{\rm host}|\pmb{\Theta}_{\rm host,blue})\Big),
\label{main_prior}
\end{align}
where $f_{\rm high\,x_{1}}$ is the fraction of the high-$x_{1}$ supernova population, $f_{\rm red|high\,x_{1}}$ is the fraction of red host galaxies in the high-$x_{1}$ supernova population,
$p_{\rm SN}$ is the prior distribution of supernova latent variables $\pmb{\phi}_{\rm SN}$, and $p_{\rm host}$ is the prior distribution of the host galaxy variables $\pmb{\phi}_{\rm host}$, 
$\pmb{\Theta}_{\rm SN}$ and $\pmb{\Theta}_{\rm host}$ are vectors of hyperparameters related to $p_{\rm SN}$ and $p_{\rm host}$. Both the supernova and host galaxy populations 
were differentiated by employing independent sets of hyperparameters, as indicated by the subscripts. Furthermore, we assumed that the only difference between high-$x_{1}$ 
supernova population in red and blue host galaxies lies in extrinsic properties captured by $E(B-V)$ and $R_{\rm B}$. Hence, all hyperparameters related to intrinsic variables 
in $\Theta_{\rm SN,high\,x_{1},red}$ and $\Theta_{\rm SN,high\,x_{1},blue}$ were kept the same (shared). We accounted for possible outliers in the host parameter space 
by including a flat background probability distribution both in the blue and red host-galaxy populations, i.e.
\begin{align}
&p_{\rm host}(\pmb{\phi}_{\rm host}|\pmb{\Theta}_{\rm host,red/blue})=(1-f_{\rm outlier})\mathcal{G}_{\rm host,red/blue}(\pmb{\phi}_{\rm host})\nonumber \\
&+f_{\rm outlier}/10,
\label{prior_host}
\end{align}
where $\mathcal{G}_{\rm host,red/blue}(\pmb{\phi})$ is a bivariate Gaussian distribution describing either the red or the blue host-galaxy population, $1/10$ is the normalisation of the second term 
given by the adopted cuts in $(g-z)$ colour ($0\le (g-z)\le 2$~mag) and the interval of the logarithmic stellar masses given by the maximum and minimum values ($5$~dex), and $f_{\rm outlier}$ 
is a free nuisance parameter shared by both host-galaxy populations.

\subsection{Likelihood}

We used the likelihood given by the probabilistic model describing the distribution of type Ia supernovae 
in a 5-dimensional space spanned by the light curve parameters $\xi_{\rm SN}$ and the host galaxy properties $\xi_{\rm host}$. 
Assuming uncorrelated measurements and data completeness expected for the volume-limited sample and minimally restrictive cuts 
in $c$ and $x_{1}$, the likelihood is given by
\begin{equation}
L \propto \prod_{i}^{N} p(\pmb{\xi_{i}}|\pmb{\Theta})=
\prod_{i}^{N}\int \mathcal{G}[\pmb{\xi}(\pmb{\phi});\pmb{\xi_{\rm obs\,i}},\mathsf{C_{\rm obs\,i}}]
p_{\rm prior}(\pmb{\phi}|\pmb{\Theta})\textrm{d}\pmb{\phi},
\label{likelihood}
\end{equation}
where $p_{\rm prior}(\pmb{\phi}|\pmb{\Theta})$ is given by eq.~(\ref{main_prior}), $\mathcal{G}[\pmb{x};\pmb{\mu},\mathsf{C}]$ 
is a multivariate Gaussian with mean $\pmb{\mu}$ and covariance matrix $\mathsf{C}$ describing the probability distribution 
of the true observable values given the measurement, $\pmb{\xi}(\pmb{\phi})$ is the vector of the measured light curve parameters 
and host galaxy properties (for $i$-th supernova), and $\mathsf{C_{\rm obs\,i}}$ is the corresponding covariance matrix with the following block structure:
\begin{equation}
\mathsf{C_{\rm obs\,i}} =
\begin{bmatrix}
\mathsf{C_{\rm SN\,i}} & 0 & 0 \\
0 & \sigma^{2}_{\log_{10}(M_{\star}/M_{\odot})\,i} & 0 \\
0 & 0 & \sigma^{2}_{(g-z)\,i}
\end{bmatrix},
\end{equation}
where $\mathsf{C_{\rm SN\,i}}$ is the full covariance matrix of the light curve parameters. Integration 
over all latent variables $\pmb{\phi}$ associated with Gaussian priors results in a multivariate normal distribution which can be factorised 
into supernova- and host-galaxy-dependent components. The remaining integration of the supernova probability component 
over $E(B-V)$ (the only variable with a non-Gaussian prior distribution) was computed numerically. The exact formulae 
are outlined in the Appendix of \citet{Wojtak2023}.

Our model contains in total 39 parameters: 15 parameters describing the prior distribution of intrinsic variables in the two supernova populations 
(mean and scatter in $M_{\rm B}$, $c_{\rm int}$ and $x_{1}$, $\alpha$, and shared $\beta$), 9 parameters corresponding to extrinsic variables 
for three supernova--host-galaxy combinations ($\tau$, mean $R_{\rm B}$ and scatter in $R_{\rm B}$ for low-stretch in red host galaxies, and  high-stretch in red/blue host galaxies), 
10 parameters describing bivariate normal distributions of the two host galaxy populations, 2 relative weights ($f_{\rm high\,x_{1}}$ and $f_{\rm blue}$), 
and 2 parameters used in the pocket effect corrections. The remaining parameter $f_{\rm outlier}$ accounts for a contribution from a flat 
probability distribution representing possible outliers in the host parameter space. 
We adopted the notation in which $\widehat{\phi}$ and $\sigma_{\phi}$ denote the mean and the dispersion of a variable $\phi$ with a Gaussian prior, i.e. $\widehat{R_{\rm B}}$ and $\sigma_{R_{\rm B}}$ for $R_{\rm B}$. For bivariate 
normal distributions describing priors of the host galaxy parameters, we used $q$ to denote the correlation coefficient. Although $\sigma_{M_{\rm B}}$ describes the scatter of $M_{\rm B}$, it is instructive to interpret this parameter as the total colour-independent unexplained scatter, which can potentially be ascribed to effects not included in the model.

\section{Results}
\label{sec:results}

We fitted the model to the data using an MCMC method implemented in the \textit{emcee} code \citep{emcee}. We used flat priors for free parameters, and the adopted 
prior limits had a negligible impact on the extent of the posterior probability distribution tails. For convergence, we required the chains to be 50 times longer than 
the largest autocorrelation time across the fitted parameters. 
The best-fit parameters are listed in Table~\ref{tab:bestmodel} as posterior means 
and uncertainties given by the 16th and 84th percentiles of the marginalised probability distributions. Figures~\ref{fig:post_int}--\ref{fig:post_ext} show the constraints on the parameters, 
grouped into those related to supernova intrinsic properties ($\{M_{\rm B},x_{1},c_{\rm int},\alpha,\beta,f_{\rm high\,x_{1}}\}$) and those related to supernova extrinsic (dust-related) 
properties ($\{E(B-V),R_{\rm B}\}$) and host galaxy properties ($\{\log_{10}(M_{\star}/M_{\odot}),(g-z)\}$).

\begin{table*}
\caption{Best-fit parameters of the prior probability distribution for the two supernova populations and their host galaxies.}
\begin{center}
\begin{tabular}{lccl}
\hline
 & low-$x_{1}$ SNe & high-$x_{1}$ SNe  & \\
\hline
$\widehat{M_{\rm B}}-5\log_{10}h_{70}\,[\rm{mag}]$ & $ -19.32 ^{+ 0.05 }_{- 0.05 }$ & $ -19.18 ^{+ 0.05 }_{- 0.05 }$ & \\
$\sigma_{M_{\rm B}}$ & $ 0.113 ^{+ 0.012 }_{- 0.011 }$ & $ 0.059 ^{+ 0.014 }_{- 0.014 }$ & \\
$\widehat{x}_{1}$ & $ -1.02 ^{+ 0.08 }_{- 0.08 }$ & $ 0.50 ^{+ 0.05 }_{- 0.05 }$ & \\
$\sigma_{x_{1}}$ & $ 0.79 ^{+ 0.05 }_{- 0.05 }$ & $ 0.55 ^{+ 0.03 }_{- 0.03 }$ & \\
$\alpha$ & $ 0.228 ^{+ 0.017 }_{- 0.017 }$ & $ 0.163 ^{+ 0.017 }_{- 0.017 }$ & \\
$\widehat{c_{\rm int}}$ & $ -0.068 ^{+ 0.007 }_{- 0.007 }$ & $ -0.080 ^{+ 0.008 }_{- 0.008 }$ & \\
$\sigma_{c_{\rm int}}$ & $ 0.029 ^{+ 0.009 }_{- 0.008 }$ & $ 0.035 ^{+ 0.007 }_{- 0.007 }$ & \\
$\beta\equiv\beta_{\rm low-x_{1}}\equiv\beta_{\rm high-x_{1}}$ & 
\multicolumn{2}{c}{$2.06 ^{+ 0.44 }_{- 0.45 }$} & \\
$f_{\rm low-x_{1}},f_{\rm high-x_{1}}\equiv 1-f_{\rm low-x_{1}}$ & $ 0.480 ^{+ 0.037 }_{- 0.037 }$ & $ 0.520 ^{+ 0.037 }_{- 0.037 }$ & \\
\hline
 &  \multicolumn{2}{l}{red, massive hosts} & blue, less massive hosts \\
  & low-$x_{1}$ SNe & high-$x_{1}$ SNe  & high-$x_{1}$ SNe \\
  \hline
$f_{\rm red|high\,x_{1}},f_{\rm blue|high\,x_{1}}\equiv 1-f_{\rm red|high\,x_{1}}$ &$-$ & $ 0.319 ^{+ 0.048 }_{- 0.048 }$ & $ 0.681 ^{+ 0.048 }_{- 0.048 }$ \\
$\widehat{R_{\rm B}}$ & $ 3.06 ^{+ 0.13 }_{- 0.13 }$ & $ 3.09 ^{+ 0.09 }_{- 0.09 }$ & $ 3.89 ^{+ 0.30 }_{- 0.29 }$ \\
$\sigma_{R_{\rm B}}$ & $ 0.62 ^{+ 0.10 }_{- 0.10 }$ & $ 0.31 ^{+ 0.07 }_{- 0.07 }$ & $ 0.64 ^{+ 0.19 }_{- 0.19 }$ \\
$\tau$ & $ 0.135 ^{+ 0.011 }_{- 0.011 }$ & $ 0.274 ^{+ 0.031 }_{- 0.031 }$ & $ 0.091 ^{+ 0.015 }_{- 0.015 }$ \\
$\widehat{\log_{10}(M_{\star}/M_{\odot})}$ & \multicolumn{2}{c}{$ 10.31 ^{+ 0.03 }_{- 0.03 }$} & $ 9.27 ^{+ 0.07 }_{- 0.07 }$ \\
$\sigma_{\log_{10}(M_{\star}/M_{\odot})}$ & \multicolumn{2}{c}{$ 0.43 ^{+ 0.02 }_{- 0.02 }$} & $ 0.72 ^{+ 0.04 }_{- 0.04 }$ \\
$\widehat{(g-z)}$ & \multicolumn{2}{c}{$ 1.18 ^{+ 0.02 }_{- 0.02 }$} & $ 0.71 ^{+ 0.02 }_{- 0.02 }$ \\
$\sigma_{g-z}$ & \multicolumn{2}{c}{$ 0.19 ^{+ 0.01 }_{- 0.01 }$} & $ 0.19 ^{+ 0.01 }_{- 0.01 }$ \\
$q$ & \multicolumn{2}{c}{$ 0.53 ^{+ 0.04 }_{- 0.04 }$} & $ 0.46 ^{+ 0.07 }_{- 0.07 }$ \\
\hline
 &  \multicolumn{3}{l}{global nuisance parameters}  \\
 \hline
 $f_{\rm outlier}$ & \multicolumn{3}{c}{$ 0.035 ^{+ 0.009 }_{- 0.009 }$}  \\
$\Delta_{x_{1}}$ (pocket effect) & \multicolumn{3}{c}{$ 0.061 ^{+ 0.045 }_{- 0.046 }$}  \\
$a_{\rm pocket}\equiv{\rm d}m_{\rm B}/{\rm d}m_{\rm B,ZTF}$ (pocket effect) & \multicolumn{3}{c}{$ 0.011 ^{+ 0.011 }_{- 0.011 }$} \\
\hline
 \end{tabular}
\label{tab:bestmodel}
\end{center}
\tablefoot{Best-fit results are summarised as the posterior mean values and errors given by credible intervals containing 68 per cent of the marginalised probabilities. The uppermost part of the table 
shows the results for supernova intrinsic latent variables ($\{M_{\rm B},c_{\rm int},x_{1},\alpha,\beta\}$), the middle part for supernova extrinsic variables ($\{E(B-V),R_{\rm B}\}$) and the global host 
galaxy properties ($\{M_{\star},g-z\}$), the bottom part for the global parameters including correction due to pocket effect. The low-stretch SN population is associated only with the red, massive host 
galaxies, whereas the high-stretch SN population contributes to both host galaxy populations.
}
\end{table*}

\subsection{Intrinsic properties}
\label{sec:res_int}

\begin{figure*}
   \centering
   \includegraphics[width=\linewidth]{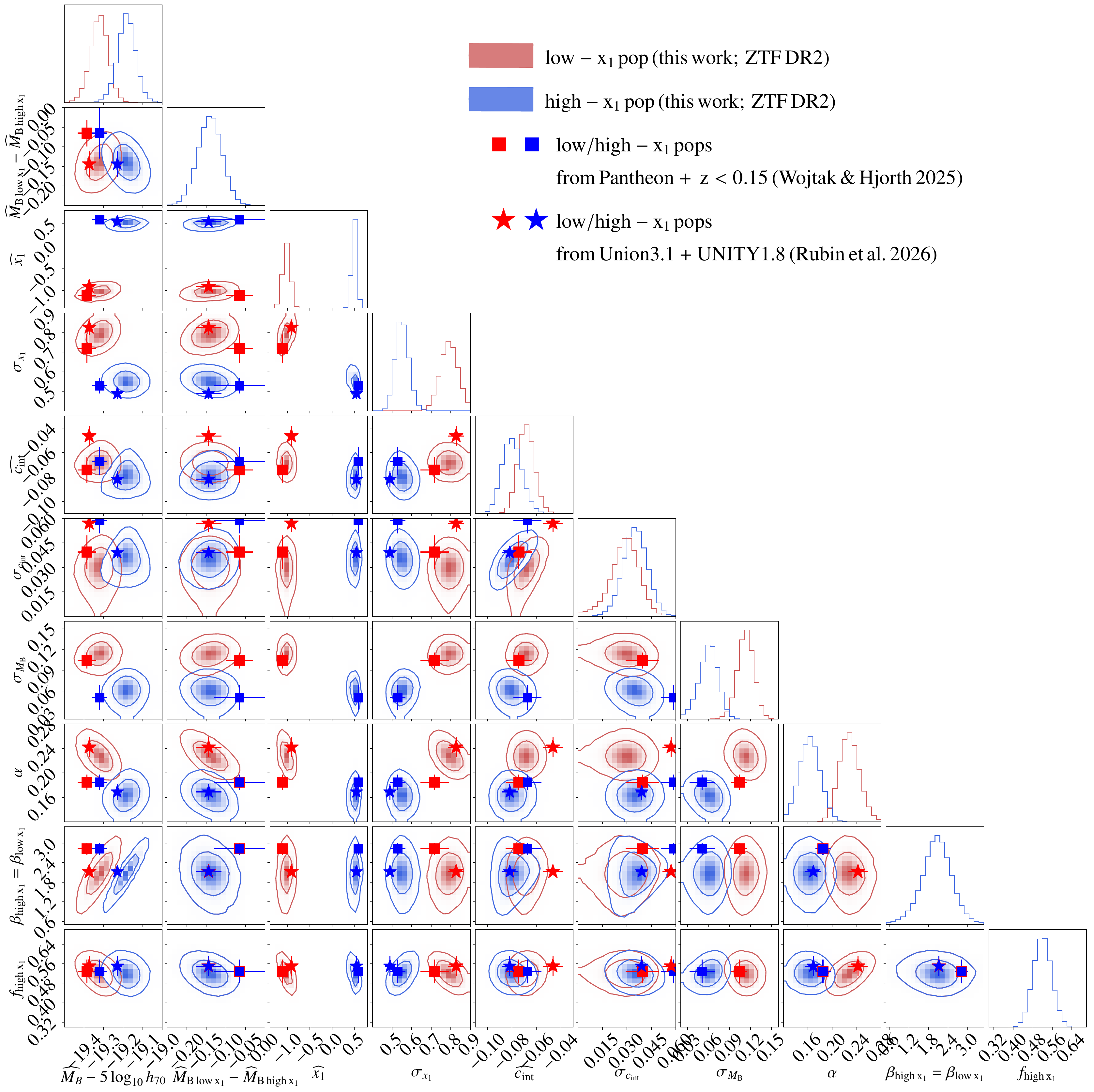}
    \caption{Constraints on the model parameters related to supernova intrinsic properties. The red and blue colours denote the low-$x_{1}$ and high-$x_{1}$ supernova populations, respectively. 
    The contours show $1\sigma$ and $2\sigma$ credible regions containing 68 and 95 per cent of the 2D marginalised probability distributions. The obtained results are compared to currently 
    available constraints on the two-population model from the Hubble-flow supernovae in the Pantheon+ catalogue \citep{Wojtak2025} and from the Union3.1 supernova compilation \citep{Rubin2026}.
    The figure demonstrates that the two supernova populations distinguished primarily by the bimodality of the stretch distribution, have different luminosities ($M_{\rm B}$), 
    stretch-correction coefficients ($\alpha$) and residual (unexplained) scatter in $M_{\rm B}$. For covariances with the host-galaxy parameters, see Figure~\ref{fig:post_full}.
    }
              \label{fig:post_int}
    \end{figure*}

\begin{figure*}
   \centering
   \includegraphics[width=\linewidth]{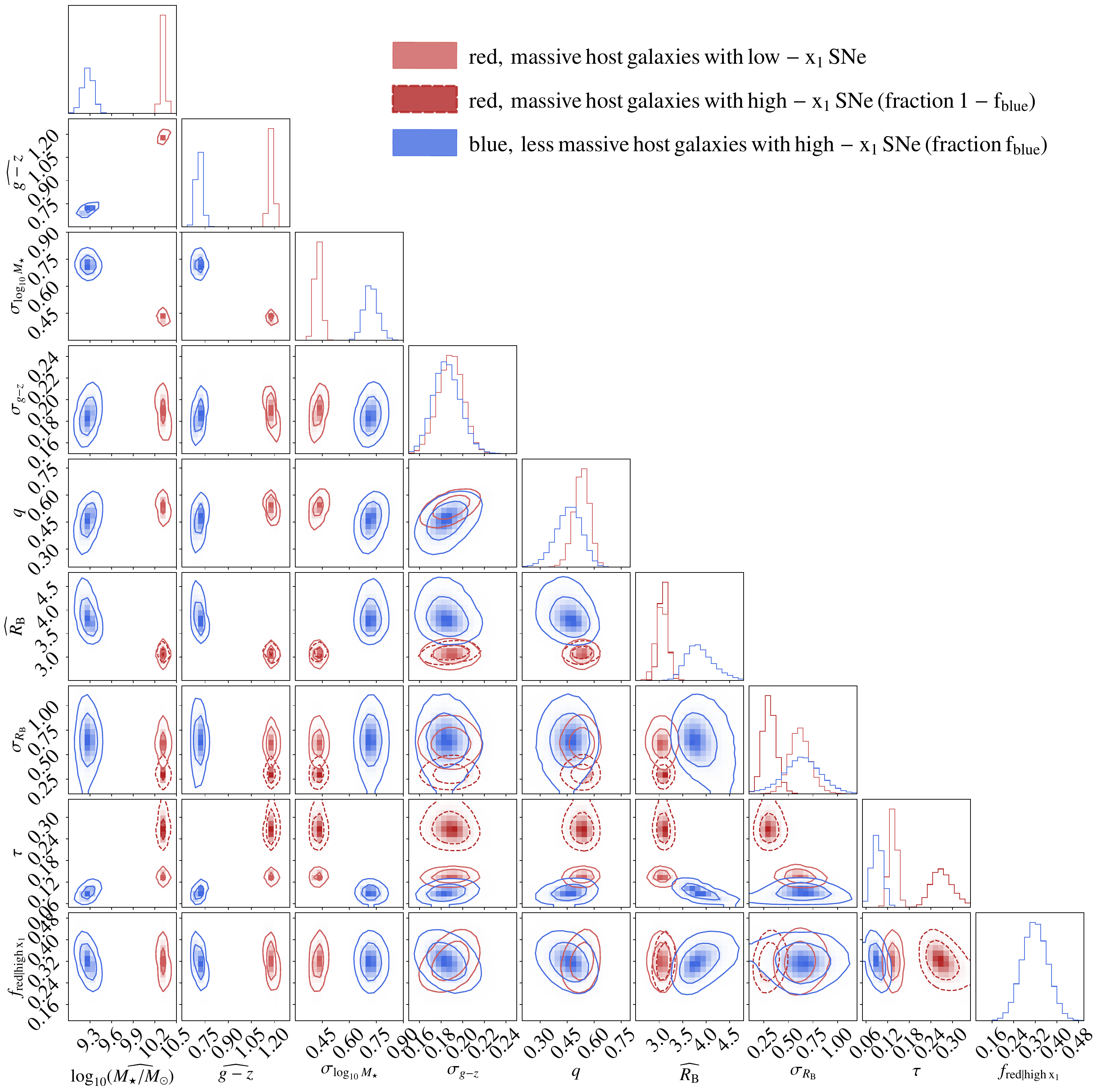}
    \caption{Constraints on the model parameters related to supernova extrinsic properties and host galaxy properties. The red and blue colours denote consistently the red/massive and 
    blue/less massive host galaxy populations. The dark red shading and contours drawn with dashed lines show the constraints on the extrinsic properties of high-stretch supernovae 
    in the red/massive host galaxies. The contours show $1\sigma$ and $2\sigma$ credible regions containing 68 and 95 per cent of the 2D marginalised probability distributions. For 
    covariances with the supernova parameters, see Figure~\ref{fig:post_full}.
    }
              \label{fig:post_ext}
    \end{figure*}

Although the stretch distribution provides most of the constraining power for distinguishing between the supernova populations, we also find a significant contribution from several other parameters. 
The supernova populations appear to have different luminosities, with low-$x_{1}$ supernovae being $\Delta M_{\rm B}=0.14\pm0.03$~mag brighter ($4.5\sigma$ significance) 
than the high-$x_{1}$ analogues (conditioned on $x_{1}=0$, $c_{\rm int}=0$, and $E(B-V)=0$). The significance of the luminosity gap is larger than what one may conclude from 
Figure~\ref{fig:post_int} due to covariance between the absolute luminosities of the two populations. The luminosity gap is the primary effect resulting from the non-linearity in the stretch correction. 
The second effect is the difference between the slopes $\alpha$, with the low-$x_{1}$ supernova population exhibiting stronger 
dependence on $x_{1}$ ($\Delta\alpha=0.065\pm0.024$; $2.7\sigma$ significance). Figure~\ref{fig:comp_x1_c} demonstrates which parts of the observational data provide constraining power on $\Delta M_{\rm B}$ and $\alpha$. The figure compares 
the mean colour-corrected magnitude as a function of $x_{1}$ to the analogous relation obtained from the best-fit model. It is apparent that the slopes of the stretch corrections are primarily 
constrained by the tails of the stretch distributions, where only one population can dominate. On the other hand, the luminosity gap is required to reproduce a very characteristic deflection 
pattern apparent between the two peaks of the stretch distribution. This non-linearity is only an apparent effect which arises from mixing the two populations with an absolute magnitude offset.

We do not find a significant difference between the supernova populations in terms of the intrinsic colour distribution, although we observe that low-$x_{1}$ supernovae are slightly redder 
($\Delta \widehat{c_{\rm int}}=0.011\pm0.011$~mag). The best-fit slope of the intrinsic colour correction is significantly flatter than the average colour correction slope of the Tripp calibration 
\citep[$\beta_{\rm T}\approx 3$ in $m_{\rm B}=M_{\rm B}+\beta_{\rm T} c-\alpha x_{1}$;][]{Ginolin2025}. However, the difference is only about $2\sigma$ significant given the current data. We tested a model assuming two independent 
$\beta$ values for both populations and found no evidence for a difference ($\beta=2.11\pm0.57$ and $\beta=2.19\pm1.08$ for the high- and low-$x_{1}$ populations, respectively).

\subsection{Host galaxy populations}

The model yields a clear separation between the derived host galaxy populations in terms of both the mean host stellar mass ($\Delta\log_{10}(\widehat{M_{\star}/M_{\odot}})=1.042\pm0.066$) 
and the mean colour ($\Delta\widehat{(g-z)}=0.467\pm0.017$~mag). The blue galaxy population exhibits nearly twice the scatter in stellar mass compared to red host galaxies. Consequently, 
there is a substantial overlap between the two populations even at masses larger than $10^{10}M_{\odot}$. The total fraction of red host galaxies derived 
from the model is $f_{\rm low-x_{1}}+f_{\rm high-x_{1}}f_{\rm red|high\,x_{1}}=0.65\pm0.03$. The contribution from the low-stretch supernova population is $74\pm4$ per cent. The obtained constraints 
on the mixture of two bivariate Gaussian distributions representing the two populations are only minimally affected by the supernova data. Fitting only the host galaxy data results in a very small 
change of the best-fit parameters with respect to the results from the joint fit including the supernova data (see Fig.~\ref{host_2D_comparison}).

\subsection{Extrinsic parameters}

We find significant differences between the scales of the $E(B-V)$ distribution for three types of supernova--host-galaxy combinations. The most extended tail of the $E(B-V)$ distribution is found 
for high-$x_{1}$ supernovae in the red host galaxies. With $\tau=0.274\pm0.031$~mag, these supernovae appear to be the primary group responsible for highly reddened supernovae 
observed in the ZTF data ($c\gtrsim 0.3$~mag). The remaining supernova populations exhibit $2$ (low-$x_{1}$ supernovae) and $3$ (high-$x_{1}$ supernovae in the blue, less massive host galaxies) 
times smaller scales of their exponential tails ascribed to the $E(B-V)$ distributions. The apparent difference between the exponential tails of high-$x_{1}$ supernovae in both host galaxy populations 
is $5.4\sigma$ significant. Figure~\ref{sn_host} shows in an intuitive way that the origin of the signal lies in the virtual lack of highly reddened supernovae ($c\gtrsim 0.4$~mag) for supernovae 
with the lowest stretch parameter ($x\lesssim -1$).

The mean extinction coefficient of high-$x_{1}$ supernovae in the blue host galaxies is consistent with a typical value of $R_{\rm B}\approx 4$ 
found in the Milky Way \citep{Fitzpatrick2007,Schlafly2016} and discy galaxies \citep{Salim2018}, which are expected to coincide closely with the underlying population of blue host galaxies. The corresponding effective colour correction for these supernovae is larger than the average colour correction slope of the Tripp calibration \citep[$\beta_{\rm T}=3$ in $m_{\rm B}=M_{\rm B}+\beta_{\rm T} c-\alpha_{\rm T} x_{1}$;][]{Tripp1998} at a $4\sigma$ significance level (estimated from an MCMC sample due to an asymmetric probability distribution). The marginalised posterior distribution also exhibits a more extended tail towards high values of $R_{\rm B}$ than in other cases, with $R_{\rm B}=4.6$ at a $2\sigma$ limit. For the red host galaxy population, the mean effective extinction coefficient found for both low-$x_{1}$ and high-$x_{1}$ supernovae is $\widehat{R_{\rm B}}=3.08\pm{0.08}$, significantly lower than $R_{\rm B}=4$ ($11\sigma$ significance combined from both populations). A relatively wide distribution of $R_{\rm B}$ with $\sigma_{R_{\rm B}}\approx 0.6$ for low-$x_{1}$ supernovae implies that explaining reddened supernovae with Milky Way-like extinction is possible for only $\lesssim 10$ per cent of the cases. On the other hand, a smaller scatter exhibited by high-$x_{1}$ supernovae rules out Milky Way-like extinction practically for every single case of reddened supernovae.

\begin{figure*}
    \centering
    \begin{subfigure}[l]{0.47\textwidth}
        \centering
        \includegraphics[width=\linewidth]{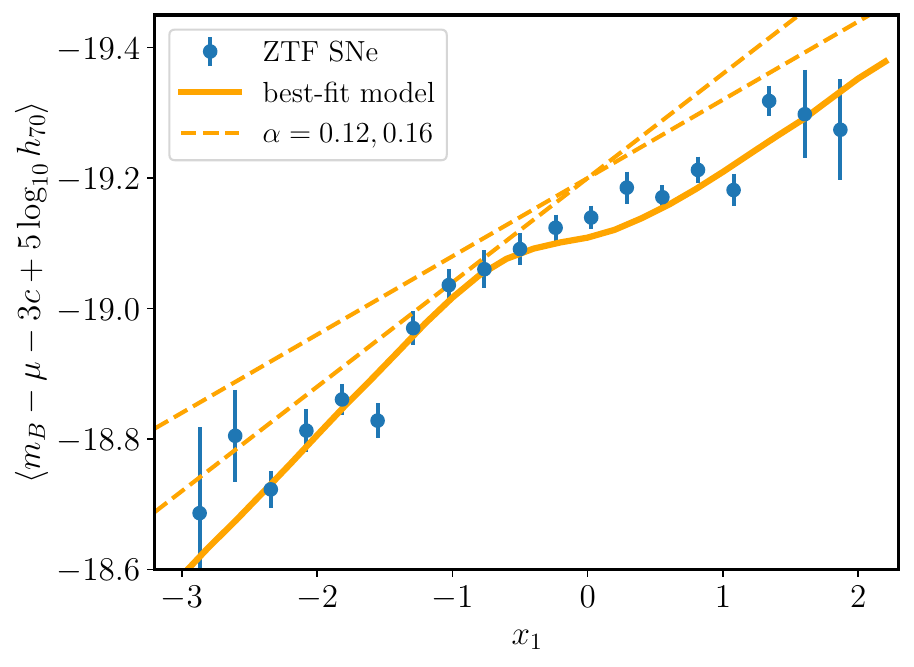} 
    \end{subfigure}
    \begin{subfigure}[r]{0.47\textwidth}
        \centering
        \includegraphics[width=\linewidth]{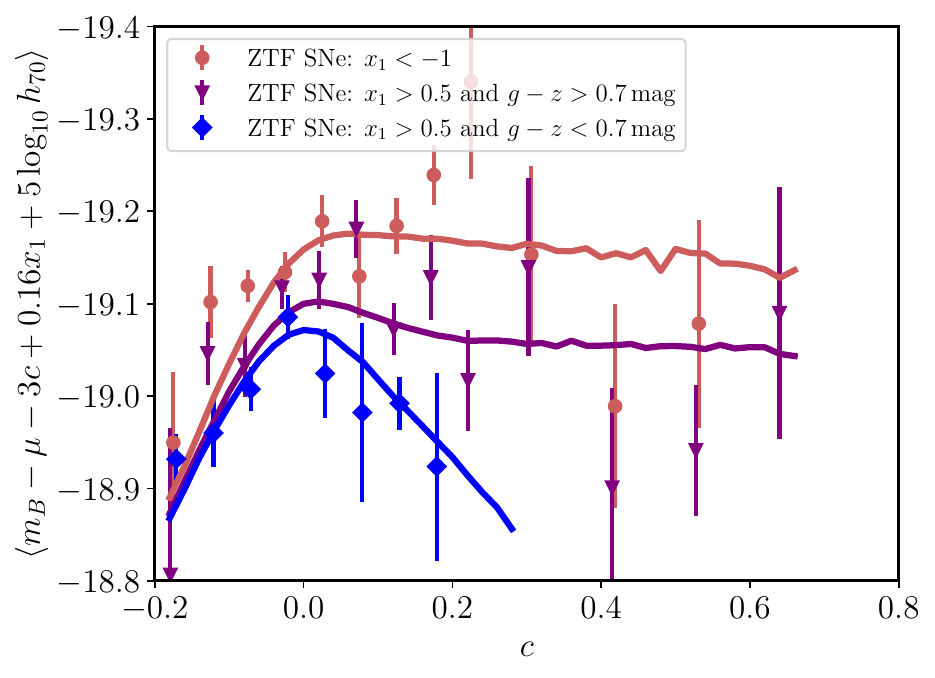} 
    \end{subfigure}
    \caption{Comparison between average observed supernova magnitude (data points) as a function of stretch (\textit{left panel}) and colour (\textit{right panel}), and the best-fit model (solid lines). The model predictions 
    were computed by sampling from the underlying prior distribution and the mean observational errors in the light curve parameters (and applying the same cuts as for the data in the right panel). The nonlinear sections of the magnitude-stretch 
    relation are reproduced by the luminosity gap between the supernova populations and the difference between the slopes. The slopes of the dashed lines span the range of typical 
    values found from fitting a linear model. The magnitude-colour relation is non-linear due to $\beta<3$ (governing the apparent slope at $c\lesssim -0.1$~mag) and $R_{\rm B}\ge 3$ 
    (governing the apparent slope at $c\gtrsim 0.1$~mag). Its shape varies across different supernova-host populations due to differences in $\widehat{R_{\rm B}}$.}
\label{fig:comp_x1_c}
\end{figure*}

The right panel of Figure~\ref{fig:comp_x1_c} demonstrates a  connection between the constrained elements of the colour correction and the data. The figure shows the mean Tripp-corrected magnitude 
as a function of colour $c$ for three subsamples representing the three cases of supernova--host-galaxy combinations. To minimise the effect of population mixing, we select supernovae with $x_{1}<-1$ 
or $x_{1}>0.5$ as representative subsamples of low- and high-$x_{1}$ populations, and galaxies with $(g-z)<0.7$~mag as a clean subsample of the blue host galaxy population. We apply the same selection 
criteria in order to compute the corresponding predictions of the best-fit model (based on Monte Carlo sampling of latent variables and adding observational noise given by average uncertainties in $c$ 
and $x_{1}$). The figure shows a one-to-one correspondence between the best-fit $R_{\rm B}$ and the apparent slopes at $c\gtrsim 0$. The apparent slope at the bluest colours ($c\lesssim -0.1$~mag) 
is reproduced by the actual slope of $m_{\rm B}-c_{\rm int}$ relation and the steepening effect resulting from binning the noisy data \citep[see e.g. Appendix A of][]{Ginolin2025}. The latter effect does not 
affect our parameter inference, which is based on a likelihood that does not involve any data binning but accounts for all errors in light curve parameters.

\subsection{Nuisance parameters}

The pocket effect is barely detected, with $\Delta_{x_{1}}=0.061\pm0.045$ and $a_{\rm pocket}=\textrm{d}m_{\rm B}/\textrm{d}m_{\rm B,ZTF}=0.011\pm0.011$. However, the obtained constraints 
agree well with estimates of the effect from simulations \citep[][]{Rigault2025} and from modelling the stretch distribution \citep[$\Delta_{x_{1}}=0.10\pm0.06$;][]{Ginolin2025}. 
We find that the impact of the pocket effect on the main parameters of our model is negligible. The two parameters most affected (maximum correlation) by $\Delta_{x_{1}}$ are $\widehat{x_{1}}$ 
(for both supernova populations), and by $a_{\rm pocket}$, $\widehat{R_{\rm B}}$ (for the low- and high-stretch populations in red host galaxies). Assuming zero values 
for both parameters of the pocket effect results 
in only a $-0.03$ shift in $\widehat{x_{1}}$ ($0.4-0.6$ of the parameter uncertainty) and a $0.014$ shift in $\widehat{R_{\rm B}}$ ($0.1-0.2$ of the parameter uncertainty).

The estimated fraction of outliers in the host galaxy parameters is about $0.035\pm0.009$. It is most likely driven by host galaxies with extremely low stellar mass estimates ($\log_{10}(M_{\star}/M_{\odot})\lesssim 7.5$). Omitting the presence of outliers in the model has a relatively small effect on the best-fit parameters: the average shift is $0.2$ of the parameter uncertainty, with a maximum of $0.7$ for the mean stellar mass of the blue galaxy population (corresponding to $-0.05$~dex shift).

\section{Discussion}
\label{sec:discussion}

We find close agreement with currently available constraints on the supernova populations based on the same approach to population modelling, 
but applied to different data sets. Figure~\ref{fig:post_int} compares our results to analogous constraints obtained by \citet{Rubin2026} from the Union3.1 
supernova compilation using the UNITY1.8 framework, and by \citet{Wojtak2025} from the Hubble-flow subset ($z<0.14$) of the Pantheon+ catalogue. 
The absolute magnitudes from the Union3.1+UNITY1.8 (defined at the mean values of $x_{1}$ in both supernova populations) were transformed to match 
absolute-magnitude definition adopted in our study (the absolute magnitude at $x_{1}=0$). The fraction of the high-$x_{1}$ population was estimated 
from its value in the lowest-redshift bin, assuming approximately equal number of host galaxies per stellar-mass bin.

Figure~\ref{fig:post_int} shows that the two Gaussian components representing the two populations are consistently constrained across supernova samples and redshifts, 
with an approximately one-to-one ratio of the two populations at low redshifts. Our study recovers remarkably well the luminosity gap between the supernova populations and 
the slopes of the stretch correction (steeper for the low-$x_{1}$ population) obtained by \citet{Rubin2026}. The luminosity gap was also confirmed in the Pantheon+ data ($z<0.15$), 
although with a somewhat smaller value ($\Delta M_{\rm B}\approx 0.065\pm0.035$~mag). Our results are qualitatively consistent with \citet{Ginolin2024a}, who found a substantially steeper 
($\Delta\alpha=0.19$) slope of the stretch correction at $x_{1}\lesssim-0.5$. However, a direct comparison between the slope estimates is not possible due to differences in the assumed 
models. In contrast to \citet{Ginolin2024a}, our model shows that part of the apparent non-linearity in the stretch correction arises from the luminosity 
gap through population mixing. Although the mass step, which was applied by \citet{Ginolin2024a}, and the luminosity gap are correlated to some degree (see below), there is no guarantee that the former 
can fully compensate for the flattening at $-0.5\lesssim x_{1}\lesssim 0.5$ in the same way as the luminosity gap between the high- and low-stretch populations.

Our analysis reveals virtually no difference between the intrinsic-colour distributions of the two populations. This is similar to the results of \citet{Wojtak2025}, but differs slightly from 
those of \citet{Rubin2026}, who found the low-$x_{1}$ population to be, on average, $0.035$~mag redder than the high-$x_{1}$ counterpart (and only $0.012$~mag redder than the mean colour 
of this population obtained in our study). The inferred slope $\beta$ of the intrinsic-colour correction is consistent with other studies modelling the effect of intrinsic colours in various 
supernova data sets \citep[see e.g.][]{Brout2021,Popovic2021,Vincenzi2024,Wojtak2025,Rubin2026}. Its signature in the ZTF data was also shown by \citet{Ginolin2025} as an apparent 
flattening of the magnitude-colour relation for the bluest supernovae.

\begin{figure*}
    \centering
    \begin{subfigure}[l]{0.47\textwidth}
        \centering
        \includegraphics[width=\linewidth]{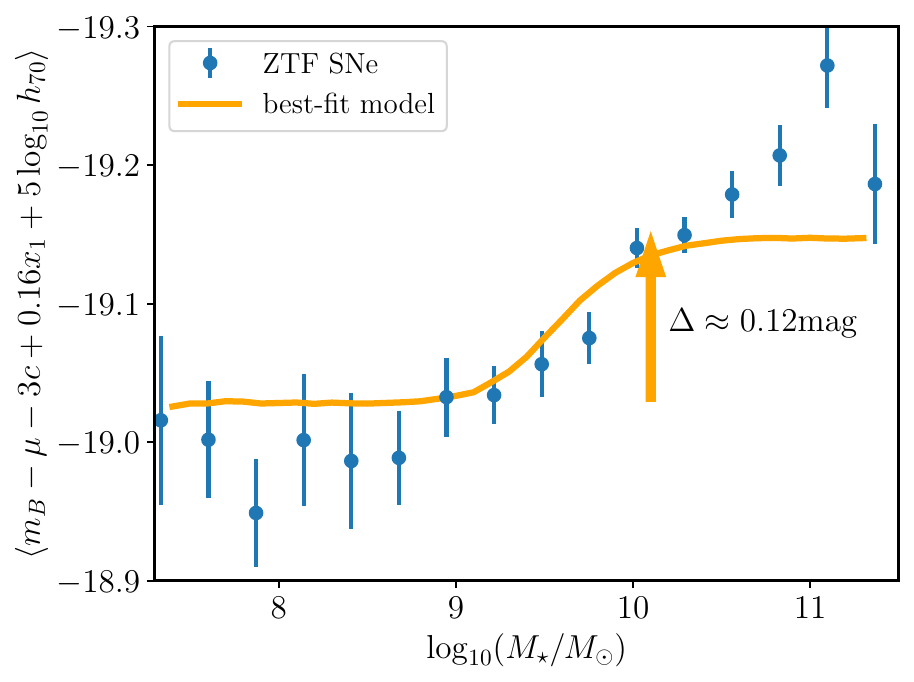} 
    \end{subfigure}
    \begin{subfigure}[r]{0.47\textwidth}
        \centering
        \includegraphics[width=\linewidth]{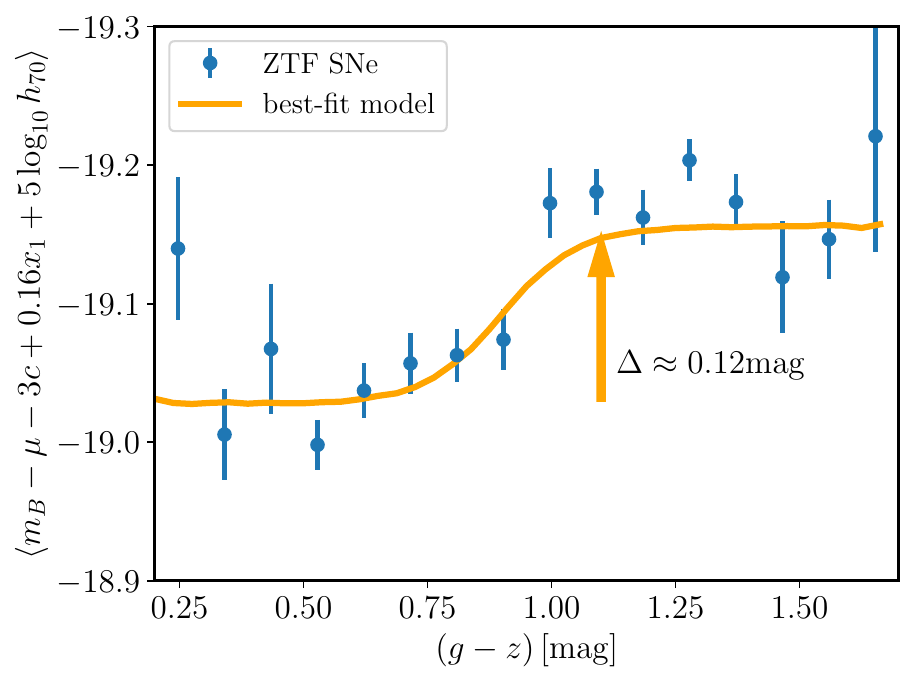} 
    \end{subfigure}
    \caption{Host-galaxy step corrections as emergent properties of the best-fit two-population model. The panels compare the mean Tripp-calibrated 
    supernovae peak luminosity as a function of stellar mass (\textit{left panel}) or rest-frame $(g-z)$ colour (\textit{right panel}) obtained from the 
    ZTF data and from the best-fit model. The model reproduces well the apparent difference between mean supernova brightness in massive (red) and 
    less massive (blue) host galaxies via different distributions of supernova populations (with different intrinsic properties) across host galaxy properties. 
    The emerging step correction (indicated by the arrows) is due to the luminosity gap between the supernova populations and different effective extinction corrections.} 
\label{step_corrections}
\end{figure*}

The mean extinction coefficient inferred from supernova magnitude-colour relation in blue host galaxies (high-$x_{1}$ population) is consistent with Milky-Way-like extinction expected in this type of galaxies. This applies to about 35 per cent of all type Ia supernovae (68 per cent of the high-$x_{1}$ population), or 42 per cent (80 per cent of the high-$x_{1}$ population) when highly reddened 
supernovae with $c>0.3$~mag are excluded. This result corroborates and generalises a range of studies finding Milky-Way-like (or sub-Milky-Way-like) extinction coefficients for host galaxies selected in 
different ways, but strongly overlapping with the blue host-galaxy population in this study, e.g. low-mass host galaxies \citep{Popovic2021,Vincenzi2024,Thorp2021,Thorp2022,Meldorf2023,Grayling2024,Rubin2026}, 
blue galaxies based on rest-frame $(u-r)$ colours \citep{Vincenzi2024}, young galaxies \citep{Wiseman2022}, or morphologically similar SH0ES calibration galaxies \citep{Wojtak2022,Wojtak2025} or nearby 
late-type galaxies \citep{Cikota2016}. The supernova colour--magnitude relation for these supernovae is strongly non-linear, and its robust separation from the remaining supernovae requires modelling both supernova and host-galaxy 
populations. The extinction coefficient inferred from low-$x_{1}$ supernovae is consistent with other measurements obtained for the corresponding population of red, massive galaxies (see the references 
above). Its low value is not compatible with extinction and the corresponding dust grain models inferred from observations. However, the inferred extinction is in fact model-dependent, and possible systematic biases cannot be ruled out. 
In this context, it is important to realise that the data constrain only the blue side of the intrinsic-colour distribution, whereas its red tail ($c_{\rm int}>\widehat{c_{\rm int}}$) is assumed to be symmetric 
through the prior construction (Gaussianity). If the red tail of the intrinsic-colour distribution were in fact more extended, and the corresponding colour correction were characterised by a similar slope $\beta$, this would naturally explain $R_{\rm B}\approx \beta\approx 2$ as an effect of intrinsic colour variation. In this picture, $R_{\rm B}\approx 3$ could arise from a combination of intrinsic colour variation and reddening due to dust extinction with $R_{\rm B}\approx 4$. Future tests may determine whether this scenario is preferred by observations. Tentative evidence can be found in \citet{Ramaiya2025}, who found a magnitude--colour relation slope of $2$ for supernovae in massive and passive host galaxies, and 
\citet{Hand2025}, who found that about 13 per cent of the variance in the phase-independent flux component of type Ia supernova light curves are not related to dust.

Although highly reddened supernovae with $0.3< c\lesssim 1$~mag have occasionally been observed in other surveys \citep{Amanullah2015,Rose2022}, the ZTF observations provided, for the first 
time, a complete sample largely unaffected by selection effects, which are particularly strong in this reddening regime. Our modelling shows that these highly reddened supernovae are primarily associated 
with the high-$x_{1}$ population hosted by red, massive host galaxies (see also Figure~\ref{sn_host}). Visual inspection of host-galaxy images for supernovae with $c>0.6$~mag reveals that this 
host-galaxy population is distinct from the predominantly early-type galaxies associated with the low-$x_{1}$ population, despite close similarities in stellar mass and colour. These hosts are primarily large, disc galaxies with strong signatures of dust (30 per cent appear edge-on, with visible dust lanes that intersect the supernova position). This points to an important role of extinction in understanding the effective slope $\widehat{R_{\rm B}}$ inferred from the data. The fact that the mean extinction coefficient inferred for this population is very similar to that of the low-$x_{1}$ population may be coincidental. The remaining two parameters relevant in this context, namely $\tau$ and $\sigma_{R_{\rm B}}$, are clearly different.

Although our model does not assume any relation between supernova absolute magnitude and host-galaxy properties, the traditional host-galaxy step corrections emerge naturally from the unequal 
distributions of supernova populations across host-galaxy properties. Figure~\ref{step_corrections} demonstrates this by comparing the data and the best-fit model in terms of the mean Tripp-calibrated 
magnitude as a function of stellar mass or rest-frame $(g-z)$ colour. The model predictions were computed by sampling all latent variables from the prior distributions, perturbing the resulting 
observables according to the mean measurement uncertainties, and applying the same Tripp corrections ($-3c+0.16x_{1}$) as used for the data. The best-fit model closely reproduces the data, and the amplitudes of the apparent step corrections are $0.116\pm0.015$~mag ($7.6\sigma$ significance) for stellar mass and $0.124\pm0.016$~mag ($7.8\sigma$ significance) for colour, where the uncertainties were estimated by sampling the model parameters from the posterior distribution. We recover both the amplitudes ($0.120-0.145$~mag) and their 
statistical significance ($5.5\sigma-6.8\sigma$) obtained by \citet{Ginolin2025} from fitting jointly host-galaxy step corrections and the Tripp calibration to the SALT2 light-curve parameters of ZTF DR2 supernovae \citep[in agreement with analogous estimates based on alternative light curve fits;][]{Ginolin2026}. This demonstrates that our model can effectively replace the standard step corrections, which are conventionally measured directly from supernova data by splitting the sample into two bins of host-galaxy stellar mass or another correlated variable. Our results match also the highest estimates of the step amplitude available from the literature, and based 
on different data sets \citep[$0.12\pm0.03$~mag and $0.09\pm0.01$~mag for the mass step correction, respectively from][]{Briday2022,Roman2018}. We can expect that alternative host-galaxy 
empirical corrections, for example with the local star formation rate or morphological type \citep[see e.g.][]{Briday2022}, can also be reproduced by our model with a comparable precision, 
through mutual correlations between host-galaxy variables. An apparent, systematic offset between the model and the data at $\log_{10}(M_{\star}/M_{\odot})>10.5$ may indicate that 
some higher-order effects are not properly accounted for. However, the offset is about 2 times smaller than the residual scatter of the low-$x_{1}$ population, which dominates in this stellar-mass range.

The host-galaxy step correction generally depends on supernova colour \citep{Vincenzi2024,Brout2021,Kelsey2021}. As shown in Figure~\ref{step_corrections_c}, this behaviour is clearly visible in the ZTF data and is well reproduced by the model. We find that approximately 40 per cent of the colour-averaged step magnitude (Figure~\ref{step_corrections}) is due to the luminosity gap between the two supernova populations, while the remaining 60 per cent arises from differences in the effective extinction corrections. The luminosity gap is therefore the primary driver of the achromatic component of the step correction. Within the framework of two-population models, the mass-step correction (or related host-galaxy corrections) appears to be a redundant term.

The apparent differences in the intrinsic properties of the two supernova populations imply two physically distinct sets of initial conditions for type Ia supernova explosions. Due to the close correlation between 
the stretch parameter and the local star formation rate \citep[see e.g.][]{Rigault2013,Rigault2020}, the initial conditions of high- and low-$x_{1}$ populations can likely be linked to two progenitor channels with single- or double-degenerate binary systems \citep{Maoz2014}. These two channels provide a natural explanation for the observed delay-time distribution \citep{Scannapieco2005,Mannucci2006,Rodney2014,Andersen2018} 
and for the diminishing contribution of the low-$x_{1}$ population at high redshifts \citep{Nicolas2021}. Further support for associating the high-$x_{1}$ population with the single-degenerate progenitor channel comes from the preferential detection of time-varying, blueshifted Na I D absorption -- possibly indicative of single-degenerate systems -- in late-type galaxies \citep[][although see also \cite{Gonzales2026}]{Maguire2013}. Furthermore, observations of bimodal nebular emission (a plausible signature of white-dwarf mergers) in early-type galaxies point to double-degenerate systems as the most likely progenitors for low-$x_{1}$ supernovae \citep{Tucker2025}. A recent study of type Ia supernova light curves from the ZTF DR2 also showed that early-time excess emission -- possibly indicating ejecta-companion interaction in single-degenerate systems -- correlates with the high-$x_{1}$ population \citep{Rojas2026}.

The lowest residual scatter in the $M_{\rm B}$ variable ($\sigma_{\rm M_{\rm B}}$) is obtained for the high-stretch population ($\sigma_{M_{\rm B}}=0.06$~mag), in agreement with the results from \citet{Wojtak2025} and \citet{Rubin2026}. From this perspective, this population can be regarded as standardisable to higher precision than low-$x_{1}$ analogues. On the other hand, the larger scatter observed for the low-$x_{1}$ population may 
indicate that some the current model is not complete.

Recent simulations of host-galaxy extinction along type Ia supernova lines of sight have shown that the expected reddening distributions are more concave than the commonly adopted 
exponential model \citep[with an excess probability at $E(B-V)\ll\tau$ and $E(B-V)\gg\tau$;][]{Hallgren2025,Duarte2026}. We tested the shape of the prior reddening distribution 
by replacing the exponential model with a Weibull distribution given by
\begin{equation}
p_{W}(E(B-V))=\frac{\gamma}{\tau}\Big(\frac{E(B-V)}{\tau}\Big)^{\gamma -1}\exp\Big(-\Big(\frac{E(B-V)}{\tau}\Big)^{\gamma}\Big).
\end{equation}
We applied this alternative model only to the high-$x_{1}$ population in blue host galaxies, for which the estimate of $\widehat{R_{\rm B}}$ clearly points to extinction as the 
dominant mechanism governing the observed supernova reddening. The Weibull distribution reduces to the exponential model when $\gamma=1$, and the simulated $E(B-V)$ 
distributions for disc galaxies are well reproduced with $0.4\gtrsim\gamma\gtrsim0.7$ \citep{Duarte2026}. Refitting the data yields $\gamma=1.01\pm0.19$. The best-fit value is fully 
consistent with the originally assumed exponential model, although the simulation-motivated distributions are not ruled out by the data.

Including host-galaxy data noticeably improves the precision of the derived supernova parameters by enabling a more effective separation of supernova populations in light-curve parameter space. 
The figure of merit (FoM) for the 28 supernova parameters is approximately $1.2\times10^4$ times larger than that obtained from a fit excluding host-galaxy data. 
This improvement is equivalent to nearly doubling the size of the supernova sample ($(\sqrt{2})^{28}\approx 1.6\times10^4$). 
Importantly, the inclusion of host-galaxy data does not lead to any significant shifts in the best-fit parameter values. The best-fit values from the joint analysis remain well within the confidence 
intervals obtained using supernova data alone. For completeness, Table~\ref{tab:bestmodelSN} summarises the constraints derived from supernova data alone and compares them with those 
obtained from the joint analysis.

\section{Summary and conclusions}
\label{sec:summary}
We modelled the light curve parameters of type Ia supernovae from the ZTF DR2 sample using a two-population mixture model. The two supernova populations are primarily defined 
by the two peaks of the stretch-parameter distribution. We employed Bayesian hierarchical modelling to constrain the relevant latent variables, such as intrinsic colours and reddening $E(B-V)$, 
and the relations between them. We augmented the model by including host-galaxy information, namely the global stellar mass and the rest-frame $(g-z)$ colour. The resulting joint modelling 
enabled us to constrain the properties of host galaxies associated with the supernova populations and improved the statistical separation of supernova environments. Our model did not 
assume any direct relation between supernova and host-galaxy variables. Instead, the apparent relations between supernovae and their host galaxies emerge from the mixing 
of distributions in the joint space of supernova light curve parameters and host-galaxy properties. Our results can be summarised as follows.
\begin{itemize}
\item The supernova populations are inferred from the ZTF data with a high level of fidelity, with their primary signature imprinted in the bimodality of the stretch-parameter distributions and, to a lesser 
extent, in statistically separable relations between the light curve parameters.
\item Including host-galaxy data noticeably improves the precision of the inferred supernova population parameters by enabling a more effective separation of the populations in light-curve parameter space, 
reducing parameter uncertainties by a factor of 2.3 on average.
\item We find strong differences in the intrinsic properties of the supernova populations, pointing to physically distinct initial conditions or progenitors channels. In addition to their distinct stretch 
distributions, the most significant differences between the populations are found in the absolute luminosity ($\Delta M_{\rm B}=0.14\pm0.03$~mag, $4.5\sigma$) and the slope of the stretch correction ($\Delta\alpha=0.065\pm0.024$, $2.7\sigma$).
\item The apparent non-linearity in the magnitude-stretch relation is fully explained by differences in luminosity and extinction between the supernova populations (manifested as an apparent flattening 
of the relation at $-1\gtrsim x_{1}\gtrsim 0.5$) and by different slopes of the relation within each population (constrained by the local slopes at $x_{1}\lesssim -1$ and $x_{1}\gtrsim 0.5$).
\item The root cause of the host-galaxy step corrections (based on stellar mass or rest-frame colour) lies in the intrinsic differences between the supernova populations, in particular their mean 
absolute magnitudes, and different extinctions. The mass-step and colour-step corrections appear as emergent properties, including the step amplitude, location, and transition sharpness, resulting from the different ways 
in which supernovae from the two populations are distributed across host-galaxy properties. The empirical mass- and colour-step corrections are well reproduced by associating the supernova populations 
with the corresponding distributions of host-galaxy variables, approximated by bivariate Gaussian distributions.
\item The mean extinction coefficient inferred for the high-$x_{1}$ population in blue, less massive host galaxies is fully consistent with the standard Milky Way extinction model ($R_{\rm B}\approx 4$). This population exhibits the strongest non-linearity in the observed magnitude-colour relation, and the mean value $\widehat{R_{\rm B}}\approx3$, which is found for the low-$x_{1}$ population, is ruled out at the $4\sigma$ significance level.
\item Supernovae from the high-$x_{1}$ population are standardisable to higher precision than their low-$x_{1}$ counterparts, with a residual scatter of only $0.06$~mag.
\item The most reddened supernovae, with $c\gtrsim 0.4$~mag, belong to the high-$x_{1}$ population associated with the red and massive host-galaxy population, 
which are predominantly disc galaxies, reddened due to strong dust attenuation. Their mean reddening is approximately 3 times larger than that of their high-$x_{1}$ 
counterparts in blue, less massive host galaxies.
\end{itemize}

Our results show that the standardisation of high-$x_{1}$ supernovae in blue host galaxies is both empirically better interpreted (through a more natural extinction correction) 
and characterised by a smaller residual scatter. From this perspective, these supernovae appear to be more robust and reliable distance indicators than the low-$x_{1}$ population. It is therefore particularly interesting to investigate whether current or future cosmological constraints can be improved by using only the high-$x_{1}$ population. Such a test can be readily implemented, at least as an additional 
consistency check: cosmological constraints derived independently from the two supernova populations are expected to be mutually consistent.

Our study also points to the relevance of two-population modelling whenever distances are inferred from different population mixtures of type Ia supernovae across redshifts. A prominent example is the Hubble constant determination based on Cepheid calibration, in which selecting galaxies with observable Cepheids automatically eliminates or strongly suppresses the low-$x_{1}$ population. Recent two-population re-modelling of supernovae from the  SH0ES-Pantheon+ data \citep{Wojtak2025} explains at least 30 per cent and up to 50 per cent of the discrepancy between the Hubble constant measured by \citet{Riess2022} and that inferred from the cosmic microwave background observation \citep{Planck2020_cosmo,Balkenhol2023}.

The evident differences in intrinsic properties of the two supernova populations (e.g. the luminosity gap) necessitate linking them to two distinct initial conditions, most likely related to 
single- and double-degenerate (prompt and delayed) progenitor channels. These and future observational constraints can provide relevant input for improving our understanding based on 
numerical simulations of type Ia supernova explosions. Incorporating progenitor age can be important for bridging the gap between initial conditions at formation and the explosion itself. Progenitor age and its effect on supernova luminosity is currently under intense investigation \citep{Yoo2026,Park2026}. The luminosity gap between the supernova populations found in this work may in fact 
correspond more closely to the progenitor-age bias, i.e. the empirical relation between supernova luminosity and age \citep{Kang2020,Lee2022,Chung2023,Chung2025}, than to the traditional 
mass-step correction.

\begin{acknowledgements}
This work was supported by research grants (VIL16599,VIL54489) from VILLUM FONDEN. RW thanks Phil Wiseman for insightful comments. The authors thank the anonymous referee for useful comments that helped improve the manuscript.
\end{acknowledgements}

\bibliographystyle{aa}
\bibliography{master}

\onecolumn
\begin{appendix}
\section{Supplementary figures and tables}

\begin{figure}[!b]
   \centering
   \includegraphics[width=0.8\linewidth]{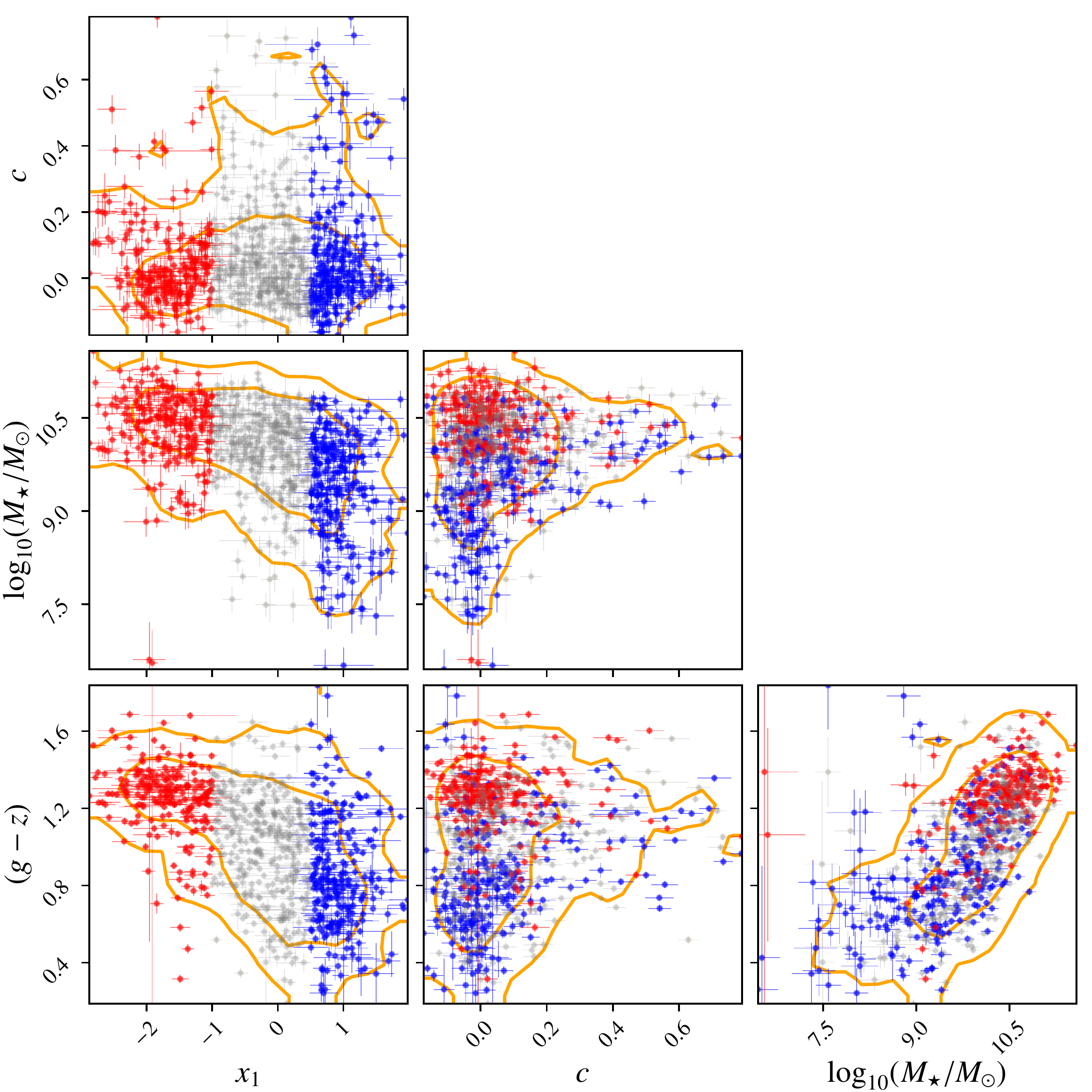}
    \caption{Correlations between the supernova stretch parameter $x_{1}$ and colour $c$, and the global stellar mass and rest-frame $(g-z)$ colour of the host galaxies. The lines show approximate iso-density contours enclosing 68 and 95 per cent of the data points in each panel. The three colours indicate approximate ranges of $x_{1}$ dominated by low-$x_{1}$ (red) and high-$x_{1}$ (blue) supernova populations, as well as an intermediate region of population mixing (see Figure~\ref{pops}). The observations show that low-$x_{1}$ supernovae are predominantly found in red ($g-z \gtrsim 1$~mag) and massive ($\log_{10}(M_{\star}/M_{\odot}) \gtrsim 10$) host galaxies, whereas high-$x_{1}$ supernovae are observed across the full range of stellar masses and colours.}
              \label{sn_host}
    \end{figure}

\begin{figure}[!b]
   \centering
   \includegraphics[width=0.6\linewidth]{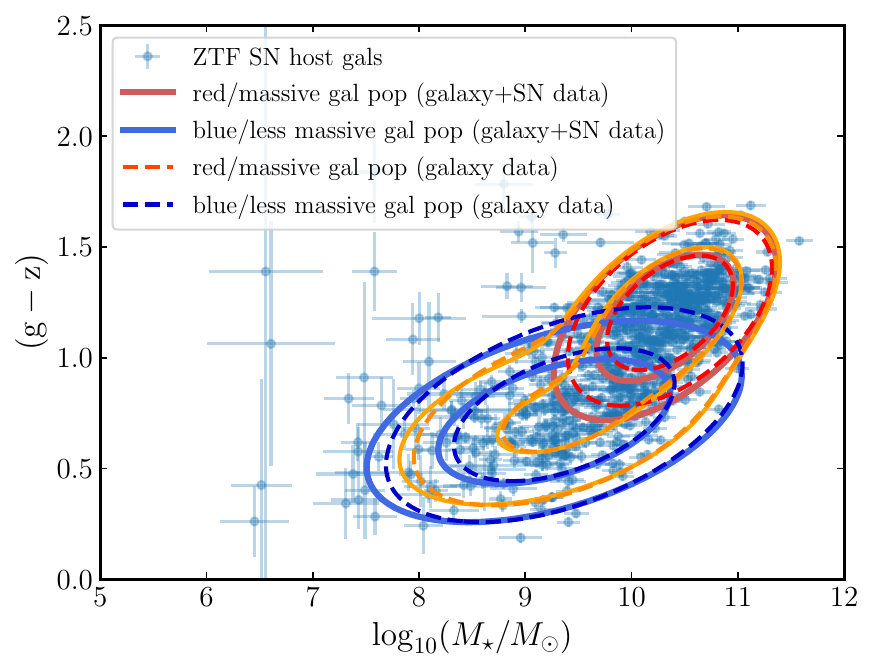}
    \caption{Comparison between the bivariate normal distributions of the two host-galaxy populations obtained from a joint analysis of supernova and host-galaxy data (solid contours) and from host-galaxy data only (dashed contours). The contours show regions containing 68 and 95 per cent of marginalised probability distributions. The orange contours show the model distribution for host galaxies from both populations.}
              \label{host_2D_comparison}
    \end{figure}    

\begin{figure*}
    \centering
    \begin{subfigure}[l]{0.47\textwidth}
        \centering
        \includegraphics[width=\linewidth]{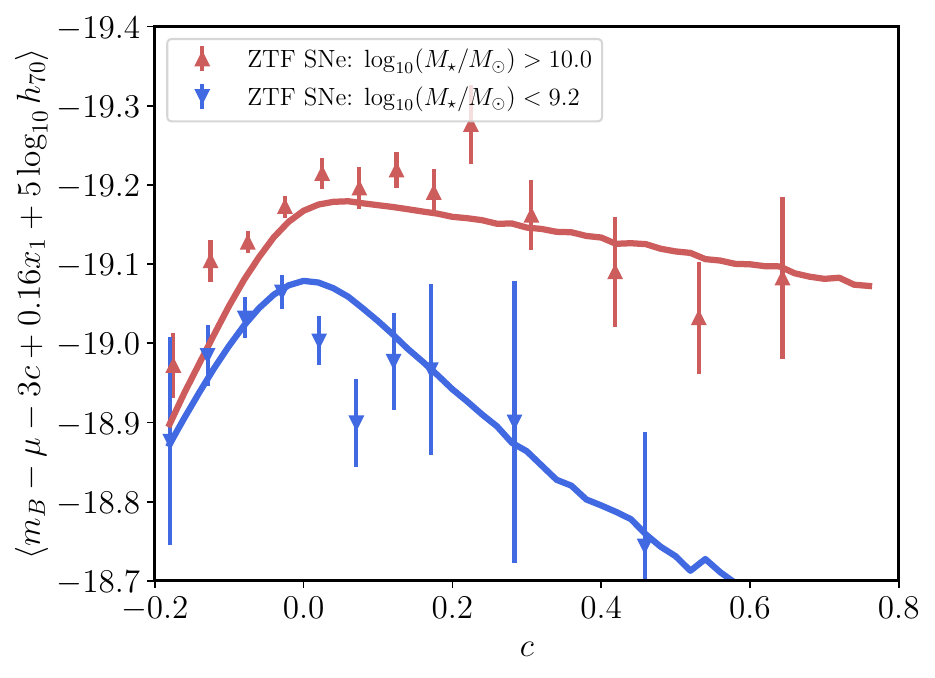} 
    \end{subfigure}
    \begin{subfigure}[r]{0.47\textwidth}
        \centering
        \includegraphics[width=\linewidth]{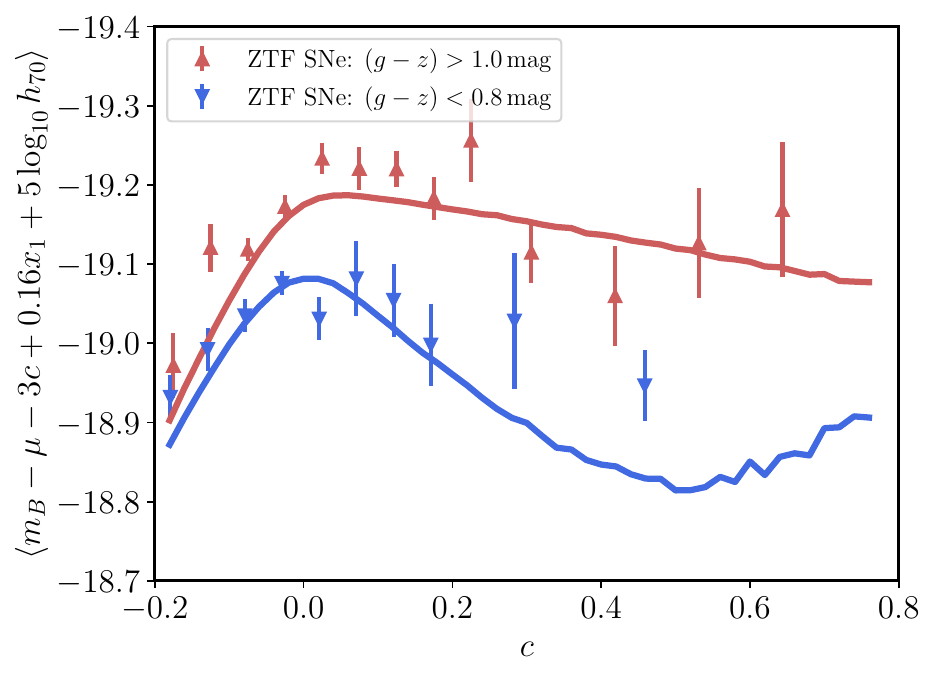} 
    \end{subfigure}
    \caption{Dependence of host-galaxy step corrections on supernova colour. The panels show 
    the mean Tripp-calibrated supernova peak luminosity as a function of supernova colour for objects below and above the apparent transition regions in the host-galaxy step corrections shown in Figure~\ref{step_corrections}. The solid lines show the model predictions, computed through Monte Carlo sampling from the model priors using the best-fit hyperparameters and perturbing the observables according to the average observational uncertainties.} 
\label{step_corrections_c}
\end{figure*}

\begin{figure*}
  \centering
  \includegraphics[width=\linewidth]{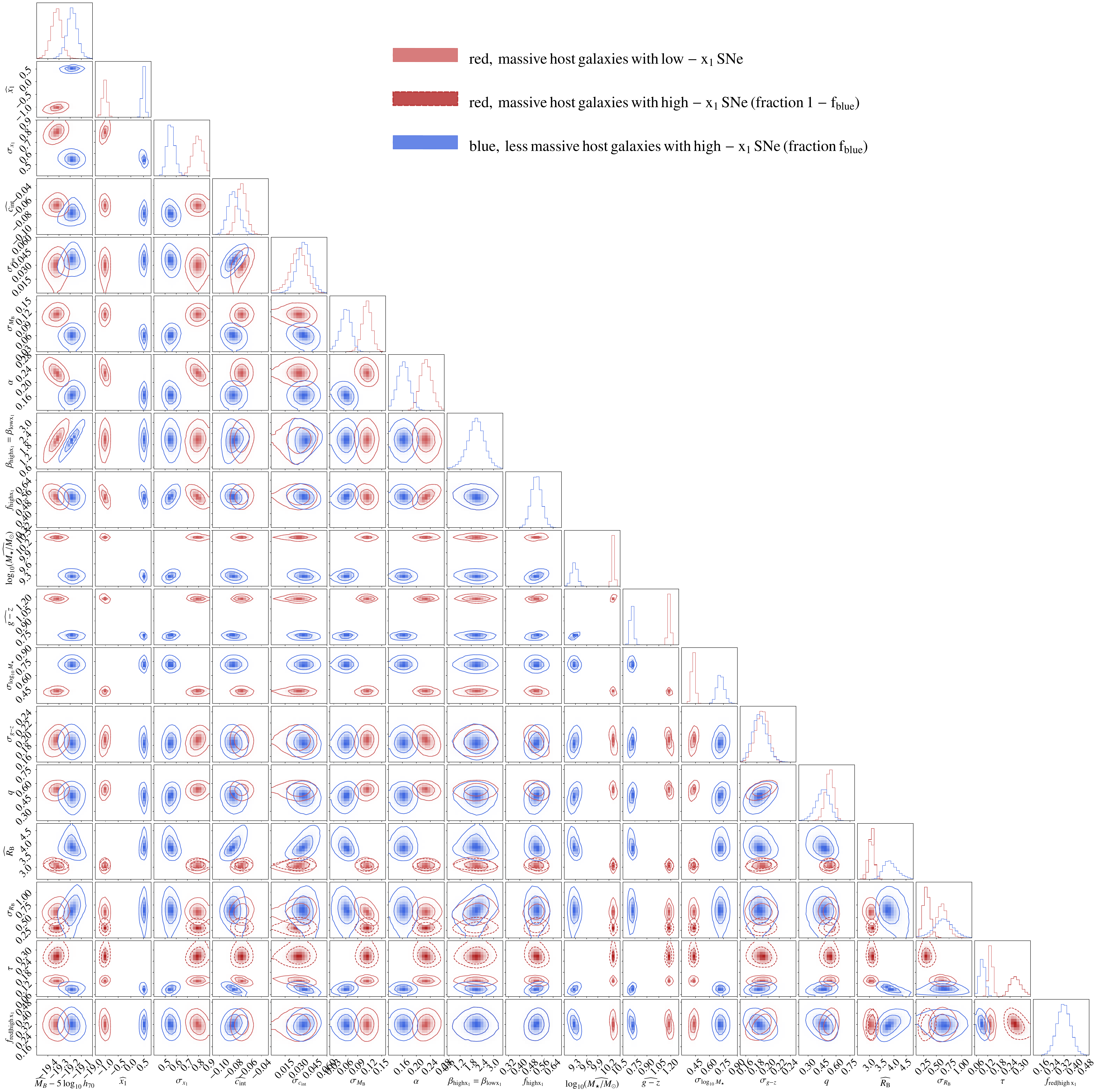}
  \caption{Constraints on all model parameters. The dark red (and dashed contours), light red and blue colours denote respectively high-$x_{1}$ supernovae in red/massive host galaxies, 
  low-$x_{1}$ supernovae in red/massive host galaxies and high-$x_{1}$ supernovae in blue/less massive host galaxies. The contours show $1\sigma$ and $2\sigma$ credible regions 
  containing 68 and 95 per cent of the 2D marginalised probability distributions.}
  \label{fig:post_full}
\end{figure*}

\begin{table*}
\caption{Best-fit parameters of the supernova prior probability distribution inferred from supernova data alone, compared with those 
obtained from the joint fit including host-galaxy data.}
\begin{center}
\begin{tabular}{lccl}
\hline
 & low-$x_{1}$ SNe & high-$x_{1}$ SNe  & \\
\hline
$\widehat{M_{\rm B}}$ & $ -19.30 ^{+ 0.06 }_{- 0.06 } ( 0.30 , 1.38 )$ & $ -19.20 ^{+ 0.06 }_{- 0.06 } ( -0.26 , 1.34 ) $ & \\
$\sigma_{M_{\rm B}}$ & $ 0.120 ^{+ 0.014 }_{- 0.013 } ( 0.53 , 1.22 )$ & $ 0.077 ^{+ 0.014 }_{- 0.014 } ( 1.19 , 1.01 ) $ & \\
$\widehat{x}_{1}$ & $ -1.11 ^{+ 0.16 }_{- 0.16 } ( -0.53 , 1.97 )$ & $ 0.49 ^{+ 0.08 }_{- 0.08 } ( -0.16 , 1.66 ) $ & \\
$\sigma_{x_{1}}$ & $ 0.74 ^{+ 0.08 }_{- 0.08 } ( -0.68 , 1.82 )$ & $ 0.55 ^{+ 0.05 }_{- 0.05 } ( -0.09 , 1.52 ) $ & \\
$\alpha$ & $ 0.219 ^{+ 0.026 }_{- 0.025 } ( -0.33 , 1.57 )$ & $ 0.122 ^{+ 0.021 }_{- 0.021 } ( -1.92 , 1.22 ) $ & \\
$\widehat{c_{\rm int}}$ & $ -0.070 ^{+ 0.007 }_{- 0.007 } ( -0.22 , 1.01 )$ & $ -0.085 ^{+ 0.010 }_{- 0.010 } ( -0.47 , 1.35 ) $ & \\
$\sigma_{c_{\rm int}}$ & $ 0.024 ^{+ 0.009 }_{- 0.010 } ( -0.52 , 1.08 )$ & $ 0.032 ^{+ 0.008 }_{- 0.008 } ( -0.37 , 1.11 ) $ & \\
$\beta$ & $ 2.04 ^{+ 0.55 }_{- 0.59 } ( -0.03 , 1.29 )$ & $ 2.04 ^{+ 0.55 }_{- 0.59 } ( -0.03 , 1.29 ) $ & \\
$f_{\rm low-x_{1}},f_{\rm high-x_{1}}\equiv 1-f_{\rm low-x_{1}}$ & $ 0.446 ^{+ 0.073 }_{- 0.072 } ( -0.46 , 1.99 )$ & $ 0.554 ^{+ 0.072 }_{- 0.073 } ( 0.46 , 1.99 ) $ & \\
\hline
 &  \multicolumn{2}{l}{red, massive hosts} & blue, less massive hosts \\
  & low-$x_{1}$ SNe & high-$x_{1}$ SNe  & high-$x_{1}$ SNe \\
  \hline
$f_{\rm red|high\,x_{1}},f_{\rm blue|high\,x_{1}}\equiv 1-f_{\rm red|high\,x_{1}}$  & $-$ & $ 0.591 ^{+ 0.192 }_{- 0.186 } ( 1.40 , 4.05 )$ & $ 0.409 ^{+ 0.186 }_{- 0.192 } ( -1.40 , 4.05 ) $ \\
$\widehat{R_{\rm B}}$ & $ 3.06 ^{+ 0.14 }_{- 0.14 } ( -0.04 , 1.13 )$ & $ 3.03 ^{+ 0.20 }_{- 0.03 } ( -0.18 , 4.33 )$ & $ 3.71 ^{+ 0.66 }_{- 0.53 } ( -0.28 , 2.12 )$ \\
$\sigma_{R_{\rm B}}$ & $ 0.65 ^{+ 0.12 }_{- 0.12 } ( 0.22 , 1.18 )$ & $ 0.36 ^{+ 0.10 }_{- 0.10 } ( 0.44 , 1.60 )$ & $ 0.93 ^{+ 0.45 }_{- 0.48 } ( 0.74 , 1.96 )$ \\
$\tau$ & $ 0.132 ^{+ 0.011 }_{- 0.011 } ( -0.26 , 1.07 )$ & $ 0.188 ^{+ 0.039 }_{- 0.023 } ( -1.52 , 1.72 )$ & $ 0.100 ^{+ 0.054 }_{- 0.043 } ( 0.17 , 3.49 )$ \\
$\Delta_{x_{1}}$ & \multicolumn{3}{c}{$ 0.081 ^{+ 0.051 }_{- 0.051 } ( 0.40 , 1.12 )$} \\
$a_{\rm pocket}\equiv{\rm d}m_{\rm B,meas}/{\rm d}m_{\rm B,true}$ & \multicolumn{3}{c}{$ 0.016 ^{+ 0.012 }_{- 0.012 } ( 0.48 , 1.01 )$} \\
\hline
 \end{tabular}
\label{tab:bestmodelSN}
\end{center}
\tablefoot{Best-fit results are summarised as the posterior mean values and errors given by credible intervals containing 68 per cent of the marginalised probabilities. 
Results obtained from the supernova-only fit (SN) and from the joint fit including host-galaxy data (SN+host) are compared 
in parentheses as: $\Big( (\rm{par}_{\rm best-fit, SN}-\rm{par}_{\rm best-fit,SN+host})/\sigma_{\rm SN},\sigma_{\rm SN}/\sigma_{\rm SN+host}\Big)$, 
where $\sigma_{\mathrm{SN}}$ and $\sigma_{\mathrm{SN+host}}$ denote the corresponding symmetrised parameter uncertainties from the two fits.
}
\end{table*}

\end{appendix}

\end{document}